\documentclass[twocolumn]{aastex631}

\usepackage{pifont}
\usepackage{threeparttable, tablefootnote}
\usepackage[version=3]{mhchem}
\usepackage{gensymb}
\usepackage{amssymb}
\usepackage{pifont}

\begin{document}

\title{Ice sublimation in the dynamic HD~100453 disk reveals a rich reservoir of inherited complex organics}

\correspondingauthor{Alice S. Booth} 
\email{alice.booth@cfa.harvard.edu}
\author[0000-0003-2014-2121]{Alice S. Booth} 
\altaffiliation{Clay Postdoctoral Fellow}
\affiliation{Center for Astrophysics \textbar\, Harvard \& Smithsonian, 60 Garden St., Cambridge, MA 02138, USA}

\author[0000-0002-7212-2416]{Lisa Wölfer}
\affiliation{Department of Earth, Atmospheric, and Planetary Sciences, Massachusetts Institute of Technology, Cambridge, MA 02139, USA}

\author[0000-0002-7935-7445]{Milou Temmink}
\affiliation{Leiden Observatory, Leiden University, 2300 RA Leiden, the Netherlands}

\author[0000-0002-0150-0125]{Jenny Calahan}
\affiliation{Center for Astrophysics \textbar\, Harvard \& Smithsonian, 60 Garden St., Cambridge, MA 02138, USA}

\author[0009-0006-1929-3896]{Lucy Evans}
\affiliation{School of Physics and Astronomy, University of Leeds, LS2 9JT, United Kingdom}
\email{l.e.evans@leeds.ac.uk}

\author[0000-0003-1413-1776]{Charles J.\ Law}
\altaffiliation{NASA Hubble Fellowship Program Sagan Fellow}
\affiliation{Department of Astronomy, University of Virginia, Charlottesville, VA 22904, USA}

\author[0000-0003-3674-7512]{Margot Leemker}
\affiliation{Dipartimento di Fisica, Università degli Studi di Milano, Via Celoria 16, 20133 Milano, Italy}

\author[0000-0003-2493-912X]{Shota Notsu}
\affiliation{Department of Earth and Planetary Science, Graduate School of Science, The University of Tokyo, 7-3-1 Hongo, Bunkyo-ku, Tokyo 113-0033, Japan}
\affiliation{Star and Planet Formation Laboratory, RIKEN Cluster for Pioneering Research, 2-1 Hirosawa, Wako, Saitama 351-0198, Japan}

\author[0000-0001-8798-1347]{Karin Öberg}
\affiliation{Center for Astrophysics \textbar\, Harvard \& Smithsonian, 60 Garden St., Cambridge, MA 02138, USA}

\author[0000-0001-6078-786X]{Catherine Walsh}
\affiliation{School of Physics and Astronomy, University of Leeds, Leeds LS2 9JT, UK}

\begin{abstract}
Protoplanetary disks around luminous young A-type stars are prime observational laboratories to determine the abundances of complex organic molecules (COMs) present during planet formation. In contrast to their lower stellar mass counterparts, these warmer disks contain the sublimation fronts of complex molecules such as \ce{CH_3OH} on spatial scales accessible with the Atacama Large Millimeter/submillimeter Array (ALMA). We present ALMA observations of the Herbig Ae disk HD 100453 that uncover a rich reservoir of COMs sublimating from the dust cavity edge. In addition to \ce{CH_3OH}, we detect \ce{^{13}CH_3OH} for the first time in a Class II disk revealing a factor of three enhancement of \ce{^{13}C} in the disk large organics. A tentative detection of \ce{CH_2DOH} is also reported resulting in a D/H of 1-2\%, which is consistent with the expected deuterium enhancement from the low temperature \ce{CH_3OH} formation in molecular clouds and with the deuteration of \ce{CH_3OH} measured in comets. The detection of methyl-formate (\ce{CH_3OCHO}), at only a few percent level of \ce{CH_3OH} is an order of magnitude lower compared to claims towards other organic-rich Herbig Ae disks but is more in line with organic abundance patterns towards the earlier stages of star formation. Together these data provide multiple lines of evidence that disks, and therefore the planet and comet-forming materials, contain inherited interstellar ices
and perhaps the strongest evidence to date that much of the interstellar organic ice composition survives the early stages of planet formation.

\end{abstract}


\section{Introduction} \label{sec:intro} 

Obtaining an inventory of the complex organic molecules (COMs; carbon and hydrogen based molecules containing 6 or more atoms) present across the different evolutionary stages of star and planet formation is fundamental when inferring the potential habitability of other worlds, and the history of the organic matter in our Solar System \citep{2009ARA&A..47..427H}. With such inventories at hand we can trace the relative abundances of different molecules across these environments to look for both similarities and differences that can be linked to the physical and chemical conditions these molecules have experienced \citep{2020ARA&A..58..727J}. The focus of this work is to determine the complex organic composition of a Class II disks with the aim
of further understanding how ices are inherited into planet-forming disks from the COMs-rich pre- and proto-stellar stages of star formation \citep[e.g.;][]{2016A&A...595A.117J, 2021MNRAS.504.5754S}, and to infer if this inherited material is transformed due to heat and/or energetic processing during incorporation into the disk or throughout the disk lifetime \citep{2014MNRAS.445..913D,2014FaDi..168..389W}.

The unique sensitivity of the Atacama Large Millimeter/submillimeter Array (ALMA) has enabled the first characterization of COMs in Class II disks \citep{2015Natur.520..198O, 2016ApJ...823L..10W,2018ApJ...862L...2F}. The initial detections of methyl-cyanide (\ce{CH_3CN}) and methanol (\ce{CH_3OH}) showed that these molecules are detectable but revealed new complications when trying to link these abundances to the potentially inherited disk ice reservoir. The abundant \ce{CH_3CN} seen towards both T-Tauri and Herbig Ae disks \citep{2018ApJ...857...69B,2018ApJ...859..131L,2021ApJS..257....9I} has been shown to trace in-situ UV driven chemistry in the disk molecular layer where the elemental carbon to oxygen ratio (C/O) is $>$ 1 \citep{2023NatAs...7...49C}. Furthermore, the \ce{CH_3OH} emission detected in the TW~Hya disk traces the non-thermal desorption of \ce{CH_3OH} ices  from the outer disk, which, is triggered by energetic particles, e.g. UV photons, and primarily results in the fragmentation of the \ce{CH_3OH} molecule upon entering the gas-phase
\citep{2016ApJ...817L..12B,2016ApJ...823L..10W,2018IAUS..332..395W}. In both cases, the rotational temperatures of \ce{CH_3CN} and \ce{CH_3OH} are lower than expected from the thermal sublimation of COMs-rich ices \citep[][Ilee et al. in prep.]{2021ApJS..257....9I} and therefore, these observations are indirect tracers of the full COMs reservoir 

In contrast, observations of transition disks around young Herbig Ae or F-type stars have revealed for the first time a thermally desorbed reservoir of COMs in Class II disks on 10's of au scales. \ce{CH_3OH} has now been detected in three such systems and is found to be tracing warm $>$100~K gas most likely within the water snowline \citep[][]{2021NatAs...5..684B, 2021A&A...651L...5V, 2023A&A...678A.146B,2024arXiv241112418T,2025arXiv250204957E}. In this region of the disk, all of the COMs should be in the gas phase and are therefore all potentially observable to us with ALMA depending on their abundances. Indeed, dimethyl-ether (\ce{CH_3OCH_3}),  methyl-formate (\ce{CH_3OCHO}) and ethelyne-oxide (\ce{c-C_2H_4O}) have also been detected in a handful of these systems \citep{2022A&A...659A..29B,2024AJ....167..164B,2024AJ....167..165B,2024ApJ...974...83Y}. Interestingly, the measured abundance ratios of these COMs relative to \ce{CH_3OH} are inferred to be at least an order of magnitude higher than that measured towards younger protostars. These initial results indicate a potential increase in the complexity of the molecular ice reservoir during the disk lifetime which could be explained via the conversion of \ce{CH_3OH} ices into larger species as seen in laboratory experiments \citep[e.g.;][]{1988Icar...76..225A, 1996A&A...312..289G, 2009A&A...504..891O}. However, there are uncertainties in determining abundances from spatially and spectrally unresolved data which may affect these results. If the underlying emitting area and/or the line optical depth of \ce{CH_3OH} are underestimated this may lead to overestimating the abundances of the other COMs \citep{2024arXiv241112418T}. One way to resolve this is via the detection of optically thinner \ce{CH_3OH} transitions and \ce{CH_3OH} isotopologues which are now routinely obtained in line surveys of younger objects \citep[e.g.;][]{2015ApJ...804...81T}. 

This paper presents ALMA observations of COMs in the disk associated with the binary system HD~100453. Located at 104~pc, the primary, HD~100453~A, is a 1.6$\mathrm{M_{\odot}}$ A9-F0 type star that hosts a dust and gas-rich transition disk \citep{2021A&A...650A.182G,2024A&A...682A.149S}. The continuum emission from the disk around HD~100453~A (hereafter, HD~100453), as viewed by ALMA, is composed of an outer dust disk with a radius from $\approx$16 to 53~au with a position angle of 145\degree~and inclination of 35\degree~and a small inner disk $<$1~au \citep{2020MNRAS.491.1335R}. VLTI/GRAVITY observations show that this inner disk is significantly misaligned with respect to the outer disk resulting in prominent shadows in the outer disk when viewed in scattered light \citep{2017A&A...597A..42B,2022A&A...658A.183B}. Beyond the outer disk of HD~100453, at $\approx$120~au, is the orbital location of the young M4 star HD~100453~B \citep{2015ApJ...813L...2W}. The dynamical interactions between the companion star and the disk drive two spiral arms traced both in scattered light and CO gas kinematics \citep{2009ApJ...697..557C, 2015ApJ...813L...2W,2018ApJ...854..130W,2020MNRAS.491.1335R}
The molecular make-up of the HD~100453 disk was surveyed by \citet{2022MNRAS.510.1148S} where \ce{C^{18}O}, \ce{HCO^+}, \ce{HCN}, CN and \ce{CS} are all detected in $\approx$1" resolution ALMA data. Notably, they report a non-detection of \ce{C_2H} hinting at a low C/O in the gas-phase. In this work we report the detections of new molecular species in the HD~100453 disk including \ce{^{13}CH_3OH} for the first time in a Class II disk. In Section 2 we describe the observations, in Section 3 we share the results and in Sections 4 and 5 discuss these results and summarize our conclusions. 

\begin{figure*}[t!]
    \centering
    \includegraphics[width=0.85\linewidth]{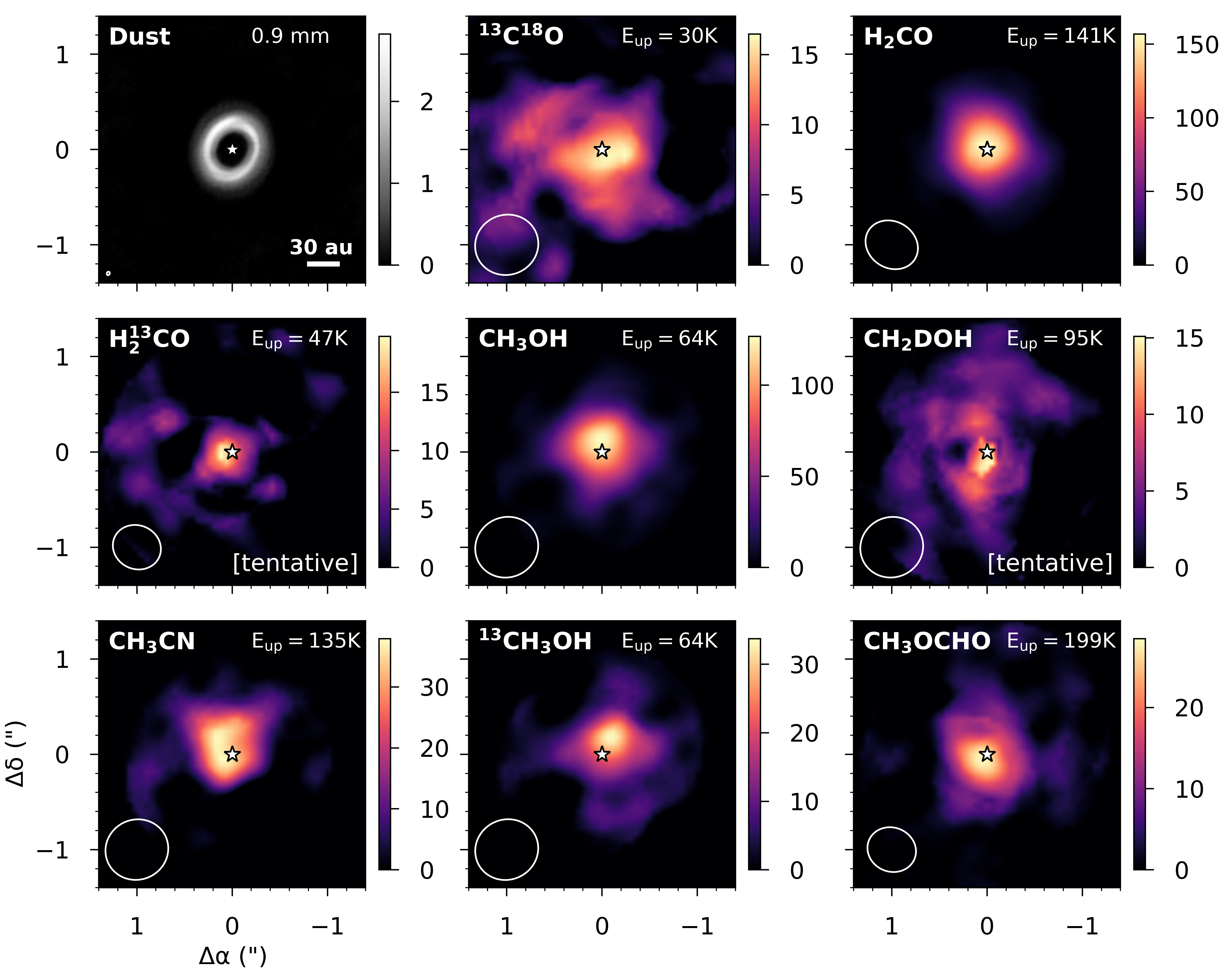}
    \caption{Integrated intensity maps of the 0.9~mm dust continuum emission (taken from \citet{2020MNRAS.491.1335R}) and molecular line emission from the HD~100453 disk.  For the line maps, the top right text notes the upper-energy level of the transition shown (see Table~\ref{table:methanol}) and in all plots the beam is shown in the bottom left corner of each panel. The units of the color bar are mJy~beam$^{-1}$~km~$\mathrm{s^{-1}}$ for the molecular lines and mJy~beam$^{-1}$ for the continuum and the position of HD~100453~A is marked with a star.}
    \label{fig:maps}
\end{figure*}

\begin{figure*}
    \centering
    \includegraphics[width=0.95\linewidth]{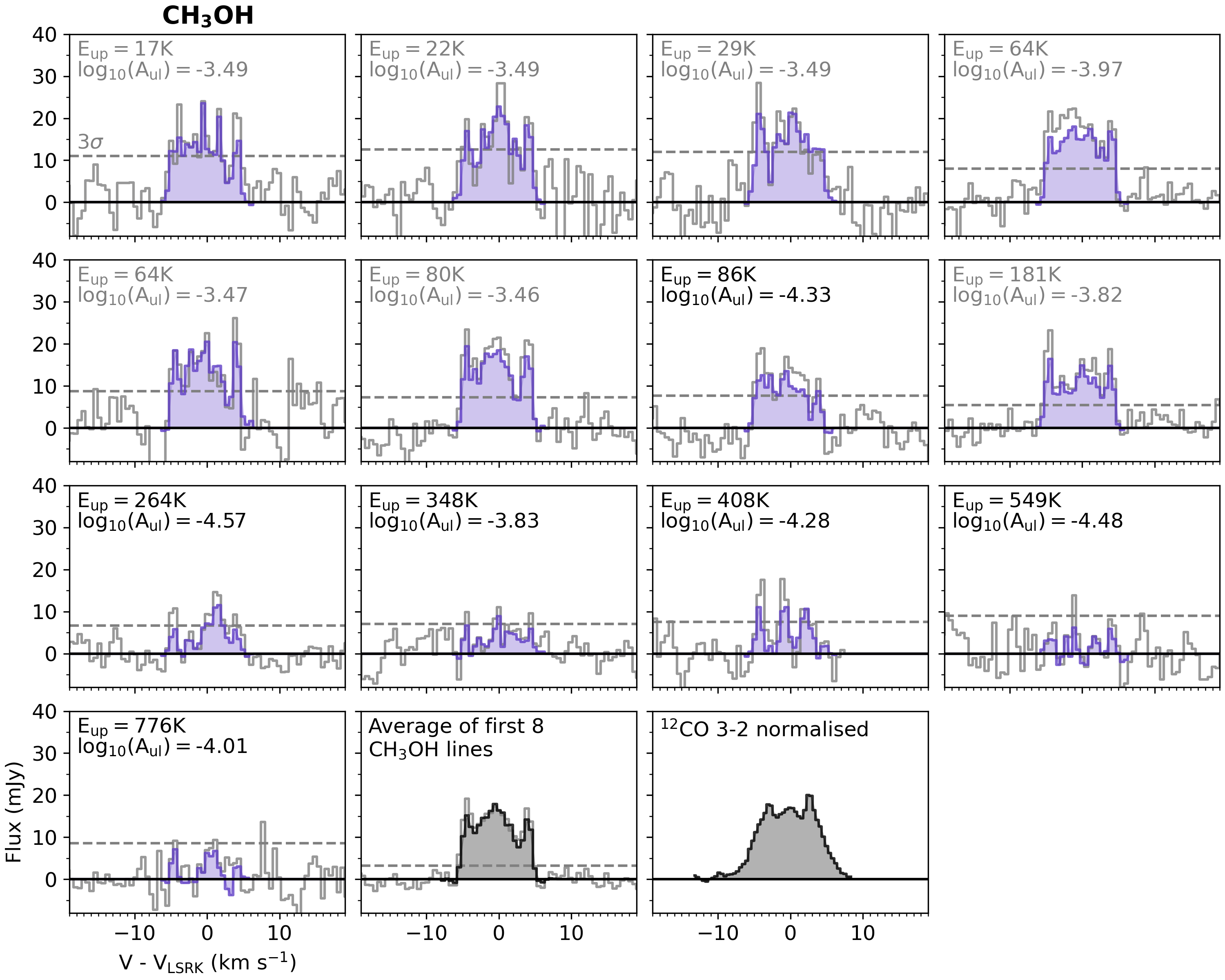}
    \caption{\ce{CH_3OH} emission line spectra extracted from a 0.5" radius circular aperture centered on-source (gray) and from a Keplerian mask with the same radial extent (purple/black). The black and gray font color for the upper energy levels and Einstein coefficients match that in Figure~\ref{fig:rotation_diagram} where only the black points are used in the rotation diagram analysis. The dashed line marks the 3~$\sigma$ level for the spectra from the circular aperture. This is an overestimate of the errors in the masked spectra but provides a good gauge of the significance of the detections. The final panel shows a normalized \ce{^{12}CO} J=3-2 disk integrated spectra for comparison where the data are first presented in \citet{2020MNRAS.491.1335R}.}
    \label{fig:ch3oh_spectra}
\end{figure*}

\section{Methods} \label{sec:observations} 

HD~100453 was observed with ALMA in Band 7 in two spectral setups within project 2023.1.00252.S (P.I. A. Booth). The first spectral setting was executed on the 13th and 14th of May 2024 for a total time of 2.6~hours with baselines ranging from 15 to 783~m and 46 antennas on both occasions. The second setting was executed on the 13th of May, 4th of September and 29th of September 2024 with the same overall baseline range as the former setting for a total time of 3.7~h with 47, 36 and 43 antennas respectively for each execution. The smallest maximal recoverable scale probed by these data are 4\farcs0~which is significantly larger than the $\approx$2\farcs diameter of the \ce{^{12}CO} gas disk \citep[e.g.;][]{2019A&A...624A..33V}. 
The two spectral set-ups are in the 290-315~GHz range and each consist of 12 spectral windows which all have a channel width of 488.281~kHz (0.51~km~s$^{-1}$ at 290~GHz). 

The data for each setting were calibrated and self-calibrated within the ALMA data reduction pipeline by ALMA staff using CASA version 6.5.5 \citep{2007ASPC..376..127M}. 
Each of the measurement sets were continuum subtracted in CASA using a fit order of 1 and all channels with clear line detections were excluded from the fit. An initial line identification was performed using matched filtering with Keplerian models \citep{Loomis2018}. These models were boolean masks tracing the expected Keplerian rotation of the disk gas based on the aforementioned position and inclination angles of the CO gas disk as traced by ALMA with a range of radial extents. 
The data were then imaged with \texttt{tCLEAN} using a Briggs robust parameter of 2.0 to maximize line detection sensitivity and a conservative analytical Keplerian mask. Images were generated with a channel width of 0.55~km~s$^{-1}$ with rms noise levels ranging from 1.0-1.6~mJy~beam$^{-1}$. 
The resulting average beam size for the images is 0\farcs$55\times$0\farcs48~(60\degree) for setting 1 and 0\farcs$67\times$0\farcs63~(-69\degree) for setting 2.

The integrated intensity maps were generated from the image cubes using a 1\farcs0 Keplerian mask\footnote{\url{https://github.com/richteague/keplerian_mask}} and the python package \texttt{BetterMoments}\footnote{\url{https://github.com/richteague/bettermoments}}. Spectra were extracted both from a 0\farcs5 radius circular mask centered at the position of HD~100453~A as well as from a Keplerian mask with the same radial extent. This mask size is $\approx$2$\times$ the beam size and was found to be sufficent to encapsulate all of line flux whilst minimizing the inclusion of noisy pixels. The disk integrated flux was then calculated from the masked spectra and for non-detections a 3~$\sigma$ upper limit on the flux was determined where $\sigma$ is propagated from the rms in the line-free channels and the number of pixels included in the mask \citep[e.g.,][]{2019A&A...623A.124C}. These fluxes are used in a now-standard rotational diagram analysis to determine the column density and rotational temperature of the gas, including an optical depth correction \citep[e.g.;][]{2018ApJ...859..131L, 2020ApJ...890..142P, 2023A&A...675A.131T}. In the case that the lines are detected in both of the spectral settings we include an additional 10\% error to account for the absolution flux calibration uncertainty of ALMA.
In the rotational diagram analysis, we used the Markov
chain Monte Carlo (MCMC) implementation of emcee-package
\citep{2013PASP..125..306F} to obtain posterior distributions of the column density and rotational temperature.
Where we do not have enough transitions for a rotational diagram approach we fix the temperature. The details on the emitting area and line width used for these calculations is discussed in Section 3. 

\begin{table*}[h]
    \centering
    \vspace*{\fill}
    \caption{Molecular data and disk integrated fluxes and upper-limits of the transitions analyzed in this work.}
    \begin{tabular}{cccccccc}
    \hline
Molecule & Transition & Frequency & log($A_{ul}$) & $E_{up}$ &  $g_{ul}$ &Beam & Integrated Flux $\pm$ Error   \\
&          & {[}GHz{]} & [s$^{-1}$]   &  [K]    &   &  ($\farcs \times \farcs$ ($^{\circ}$))   & [mJy km s$^{-1}$]       \\ \hline \hline

\ce{^{13}C^{18}O} & 3-2    & 314.119653     &-5.723 & 30.2  & 7  & 0.67$\times$0.63 (-69.0)   & 20.2$\pm$5.6\\ \hline 

\ce{HDO} & 6(2,5)- 5(3,2)  & 313.750620      & -4.426 & 553.7 &  13  & 0.66$\times$0.63 (-67.0) &  $<$16.8\\  \hline

\ce{H_2CO} & 4(1,3)-3(1,2) & 300.836636    & -3.144  & 47.8 & 27  & 0.38$\times$0.32 (-51.0)  & 340.7$\pm$17.1\\  
\ce{H_2CO} & 4(2,2)-3(2,1) & 291.948067    & -3.280 & 82.1 &  9  & 0.54$\times$0.46 (65.0)  & 122.0$\pm$6.4 \\  
\ce{H_2CO} & 4(3,2)-3(3,1) & 291.380442    & -3.517 & 140.9  & 27 & 0.57$\times$0.49 (58.0)& 165.7$\pm$7.0$^{*}$\\  
\ce{H_2CO} & 4(3,1)-3(3,0) & 291.384362    &-3.517  & 140.9  & 27 & 0.57$\times$0.49 (58.0) & 165.7$\pm$7.0$^{*}$\\  \hline

\ce{H_2^{13}CO} & 4(1,3)-3(1,2) & 293.126515 &-3.178& 47.0&  27  & 0.51$\times$0.46 (66.0) &  16.1$\pm$5.5\\  \hline

\ce{CH_3OH} & 1(1,0)-1(0,1)     & 303.366921 & -3.493 & 16.9  & 12  & 0.55$\times$0.47 (57.0)& 129.1$\pm$8.5\\
\ce{CH_3OH} & 2(1,1)-2(0,2)     & 304.208348 & -3.490 & 21.6  & 20  & 0.52$\times$0.45 (64.0)& 148.7$\pm$8.9\\
\ce{CH_3OH} & 3(1,2)-3(0,3)     & 305.473491 & -3.486 & 28.6  & 28  & 0.49$\times$0.44 (68.0)& 147.3$\pm$8.8\\
\ce{CH_3OH} & 6(1,5)-5(1,4)     & 292.672889 & -3.975 & 63.7  & 52  & 0.67$\times$0.63 (-64.0)& 143.0$\pm$8.4\\
\ce{CH_3OH} & 6(1,5)-6(0,6)     & 311.852612 & -3.466 & 63.7  & 52  & 0.57$\times$0.48 (58.0)& 140.6$\pm$8.4\\
\ce{CH_3OH} & 7(1,6)-7(0,7)     & 314.859528 & -3.456 & 80.1  & 60  & 0.66$\times$0.62 (-69.0)& 140.2$\pm$7.2\\
\ce{CH_3OH} & 7(-1,6)-6(-2,5)   & 313.596760 & -4.327 & 86.0  & 60  & 0.66$\times$0.63 (-67.0)& 76.2$\pm$7.8\\
\ce{CH_3OH} & 12(0,12)-11(1,11) & 302.912979 & -3.815 & 180.9 & 100  & 0.68$\times$0.64 (-63.0)& 107.6$\pm$5.8\\
\ce{CH_3OH} & 10(-5,5)-11(-4,7) & 302.830740 & -4.569 & 263.7 & 84  & 0.68$\times$0.64 (-63.0)&  46.6$\pm$4.8\\
\ce{CH_3OH} & 4(1,4)- 5(2,4) & 312.247362 & -3.825 & 348.4 & 36  & 0.67$\times$0.63 (-63.0)& 37.9$\pm$4.6\\
\ce{CH_3OH} & 18(1,18)-17(2,15) & 306.291135 & -4.283 & 407.6 & 148  & 0.49$\times$0.44 (68.0) & 47.3$\pm$6.1\\
\ce{CH_3OH} & 17(6,11)-18(5,14) & 291.908215 & -4.482 & 548.6 & 140  & 0.57$\times$0.48 (58.0) & 14.8$\pm$7.3\\
\ce{CH_3OH} & 12(-8,5)-12(-7,5) & 294.256085 & -4.009 & 775.7 & 100  & 0.51$\times$0.46 (67.0) & 19.2$\pm$7.2\\ \hline

\ce{^{13}CH_3OH} &3(1,2)-3(0,3) & 305.699456 & -3.485 & 28.3 & 7   & 0.49$\times$0.44 (70.0) &  30.1$\pm$6.6\\  
\ce{^{13}CH_3OH} &6(1,5)-6(0,6) & 311.773919 & -3.466 & 62.5 & 13  & 0.67$\times$0.63 (-64.0) & 31.9$\pm$5.4 \\  
\ce{^{13}CH_3OH} &7(1,6)-7(0,7) & 314.635901 & -3.457 & 78.5 & 15  & 0.66$\times$0.62 (-69.0) & 35.4$\pm$5.6 \\  
\ce{^{13}CH_3OH} &10(0,10)-9(1,8) & 302.590285 & -4.160 & 137.5 & 21  & 0.68$\times$0.64 (-63.0) & 11.6$\pm$4.3 \\  
\hline

\ce{CH_2DOH} & 7(2,6)-6(2,5) o1 &  312.324908 & -3.915 & 95.5  & 15 & 0.66$\times$0.63 (-67.0)&  21.2$\pm$4.9$^{*}$ \\ 
\ce{CH_2DOH} & 7(2,6)-6(2,5) e1 &  312.322940 & -3.932 & 86.5  & 15 & 0.66$\times$0.63 (-67.0)&  21.2$\pm$4.9$^{*}$  \\  \hline

\ce{CH_3CN} & 17(0)-16(0) & 312.687743 & -2.737 & 135.1 & 70   & 0.66$\times$0.63 (-67.0)& 42.9$\pm$5.7\\ 
\ce{CH_3CN} & 17(1)-16(1) & 312.681731 & -2.739 & 142.2 & 70   & 0.66$\times$0.63 (-67.0)& 35.1$\pm$5.4\\  
\ce{CH_3CN} & 17(2)-16(2)  & 312.663700 & -2.743 & 163.6 & 70  & 0.66$\times$0.63 (-67.0) & 27.2$\pm$6.0\\  
\ce{CH_3CN} & 17(3)-16(3)  &312.633653 & -2.751 & 199.4 & 140  & 0.66$\times$0.63 (-67.0) & 21.0$\pm$5.4\\  \hline

\ce{CH_3OCHO} & 27(1,27)-26(1,26) A & 291.111871 & -3.428&  198.6  &110 & 0.52$\times$0.46 (68.0)&  38.4$\pm$5.1$^{*}$\\ 
\ce{CH_3OCHO} & 27(0,27)-26(0,26) A & 291.111891  & -3.428&  198.6& 110   & 0.52$\times$0.46 (68.0)& 38.4$\pm$5.1$^{*}$ \\  
\ce{CH_3OCHO} & 27(1,27)-26(1,26) E & 291.111192  & -3.428 &  198.6  & 110 & 0.52$\times$0.46 (68.0)&   38.4$\pm$5.1$^{*}$\\  
\ce{CH_3OCHO} & 27(0,27)-26(0,26) E & 291.111212 & -3.428&   198.6 &  110 & 0.52$\times$0.46 (68.0) & 38.4$\pm$5.1$^{*}$\\   \hline


\ce{CH_3OCH_3} & 16(1,16)-15(0,15) EA & 292.412244 & -3.705 &120.3 &  198  & 0.51$\times$0.46 (66.0&  $<$22.0$^{*}$\\  
\ce{CH_3OCH_3} & 16(1,16)-15(0,15) EA & 292.412244 &-3.706 &120.3 &  132  & 0.51$\times$0.46 (66.0&  $<$22.0$^{*}$\\  
\ce{CH_3OCH_3}& 16(1,16)-15(0,15) EE & 292.412416 & -3.705 &120.3 &  528  & 0.51$\times$0.46 (66.0&  $<$22.0$^{*}$\\
\ce{CH_3OCH_3} & 16(1,16)-15(0,15) AA & 292.412588 & -3.705 &120.3 &  330  & 0.51$\times$0.46 (66.0&  $<$22.0$^{*}$\\ 

    \hline
    \label{table:methanol}
    \end{tabular}
    \tablecomments{All molecular data are taken from The Cologne Database for Molecular Spectroscopy \citep[CDMS][]{2001A&A...370L..49M,2005JMoSt.742..215M} where available for consistency. The data for \ce{CH_3OCHO}, HDO, \ce{CH_2DOH} are from the Molecular Spectroscopy and Line Catalogs collated by the NASA Jet Propulsion Laboratory
 \citep{1998JQSRT..60..883P}. \newline $^{*}$~notes a blended transition of the same molecule where the flux reported is the sum of both components.}
\end{table*}

\section{Results} \label{sec:results} 

\subsection{Emission morphology and kinematics}

In total we detect five organic molecules in the HD~100453 disk above the 5~$\sigma$ level in the disk integrated fluxes which are extracted from Keplerian masks. These molecules are \ce{H_2CO}, \ce{CH_3OH}, \ce{^{13}CH_3OH}, \ce{CH_3CN} and \ce{CH_3OCHO}. In addition, we also detect the rare CO isotopologue \ce{^{13}C^{18}O}.
We also identify \ce{CH_2DOH} at the 4~$\sigma$ level in the disk integrated line flux and \ce{H_2^{13}CO} at the 3~$\sigma$ level. The list of different transitions detected for each molecule are shown in Table~\ref{table:methanol} along with the associated disk integrated lines fluxes and errors. Within these data we note the detection of the \ce{H_2CO} J=4(2,3)-3(2,2) and \ce{CH_3OH} J=15(1,14)-14(2,13) transitions but these are blended in velocity space and are therefore excluded from the analysis. In addition, the \ce{CH_3OH} J=17(6,11)-18(5,14) line is only detected at a 2~$\sigma$ significance level but as it follows the same line shape as the other transitions we include this data point, and the associated error, in our subsequent analysis. In Table~\ref{table:methanol}, we also list a select number of transitions that went undetected for molecules of particular interest HDO and \ce{CH_3OCH_3} and their associated upper-limits. 

Figure~\ref{fig:maps} presents the Keplerian masked integrated intensity maps of a sub-set of the detected lines in the HD~100453 disk alongside the higher angular resolution 0.9~mm continuum observations from \citet{2020MNRAS.491.1335R}. All of the molecular emission is compact with respect to the beam and appear to be co-spatial. Of note, are the J=6(1,5)-6(0,6) ($\mathrm{E_{up}=64~K}$) transitions of \ce{CH_3OH} and \ce{^{13}CH_3OH} which show only a factor $\approx$3.3 difference in peak values in their respective integrated intensity maps. This low value when compared to the \ce{^{12}C}/\ce{^{13}C} ratio of $\approx$69 expected for the local interstellar medium likely indicates optically thick emission from the main \ce{CH_3OH} isotopologue for this transition \citep{1999RPPh...62..143W}.

The extracted spectra of the total 13 isolated \ce{CH_3OH} lines which cover a range in upper energy levels from 17 to 776~K are shown in Figure~2. 
In addition to \ce{CH_3OH} we also detect four lines of \ce{^{13}CH_3OH} and these spectra are shown in Figure~3 alongside the spectra for \ce{CH_3CN}, \ce{CH_3OCHO} and the other tentative and non-detections of note. Given the large cavity in mm-dust emission we would expect ringed emission from these species although we do not have the spatial resolution to resolve this in the image plane. To increase the relative signal-to-noise in the \ce{CH_3OH} spectra we also show the average spectra calculated from the first eight transitions listed in Table~\ref{fig:ch3oh_spectra}. This was generated by re-gridding and stacking 8 individual measurement sets that were split out around each of the lines. These steps were performed using the CASA tasks \texttt{mstransform} and \texttt{concat}. 
These measurement sets were then concatenated and imaged and the spectra were extracted following the same procedure as for the individual line images. The \ce{CH_3OH} line wings drop off sharply at approximately -5.5~km~s$^{-1}$ and +5.0~km~s$^{-1}$ relative to the source velocity of 5.25~km~s$^{-1}$. The difference in the red and blue wings of the line may indicate an underlying asymmetry in the gas emission but it is within one velocity channel of the data. If we assume an inclination angle of 35$\degree$ and a stellar mass of 1.6~M$_{\odot}$ the average velocity of the line wings is consistent with a radial distance of $\approx$17~au. Therefore, these line wings are tracing gas at the mm-dust cavity edge. 

The lines also show an interesting morphology with three emission peaks compared to the expected double peaked profile associated with the Keplerian rotation of a ring of gas. 
To place this in context with the bulk gas of the disk we also show in Figure~2 the total disk integrated spectra of the \ce{^{12}CO} J=3-2 line,  here these data are presented fully in \citet{2020MNRAS.491.1335R} and \citet{2023A&A...670A.154W}. Compared to the \ce{CH_3OH}, the CO line wings trace higher velocities and therefore tracing gas closer to the star, within the dust cavity, but interestingly there is a similar bump in the spectra around the source velocity as seen in the \ce{CH_3OH}.
To investigate the morphology of the \ce{CH_3OH} emission further we imaged the stacked data at lower robust values of +0.5,-0.5 and -2.0. The resulting integrated intensity maps, peak emission maps and peak velocity maps are shown in Figure~\ref{fig:4} and the properties of the images in Table~\ref{stacked_image_properties}. When going to higher resolution the emission becomes more compact and there is a clear offset in the integrated intensity to the North East of the disk. This excess on the minor axis may reflect the excess emission at the source velocity highlighted in the spectra.

\begin{figure*}
    \centering
    \includegraphics[width=0.95\linewidth]{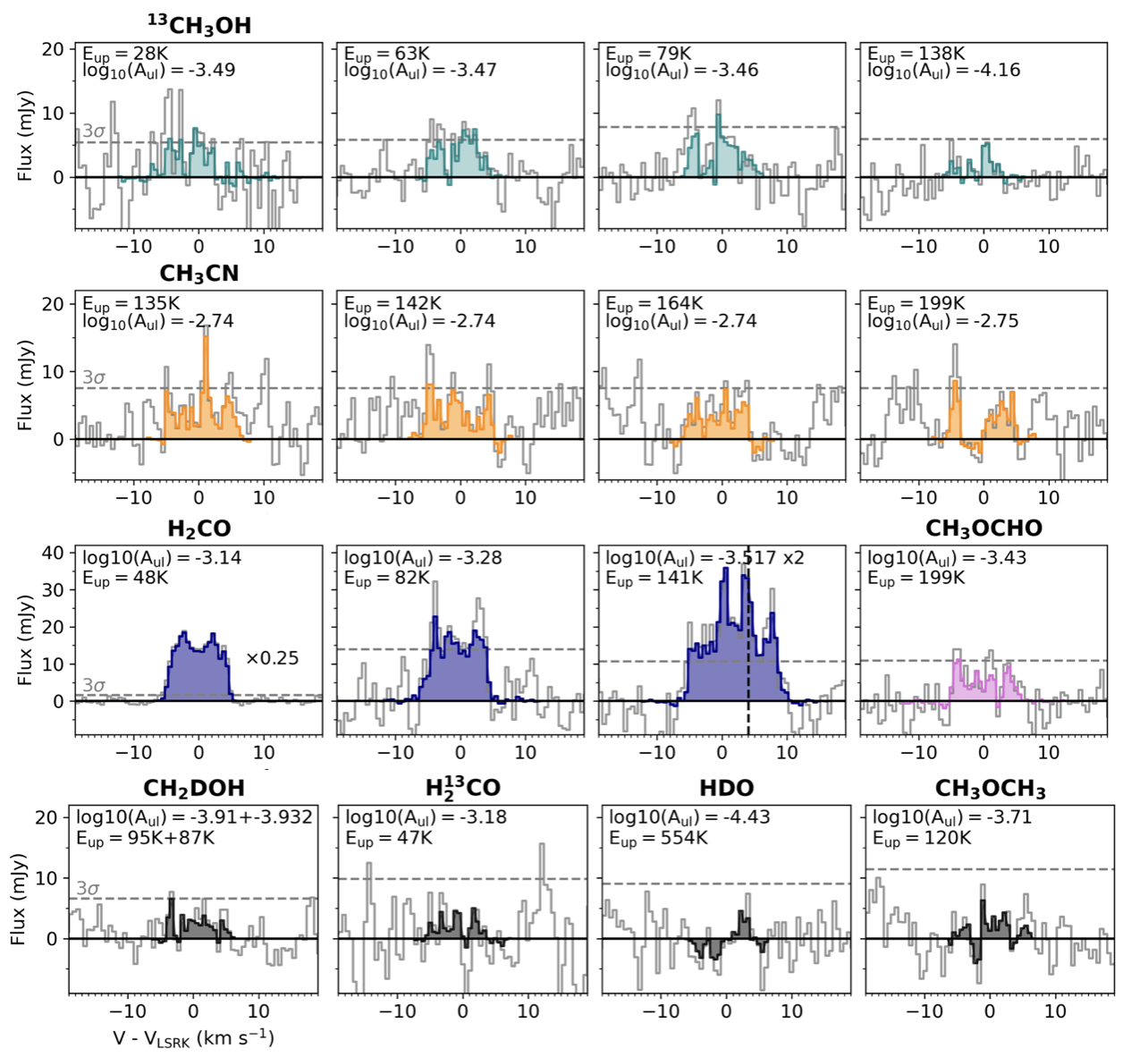}
    \caption{Same as Figure~\ref{fig:ch3oh_spectra} for the other molecular lines in the data.}
    \label{fig:other_spectra}
\end{figure*}

\begin{figure*}
    \centering
    \includegraphics[width=0.95\linewidth]{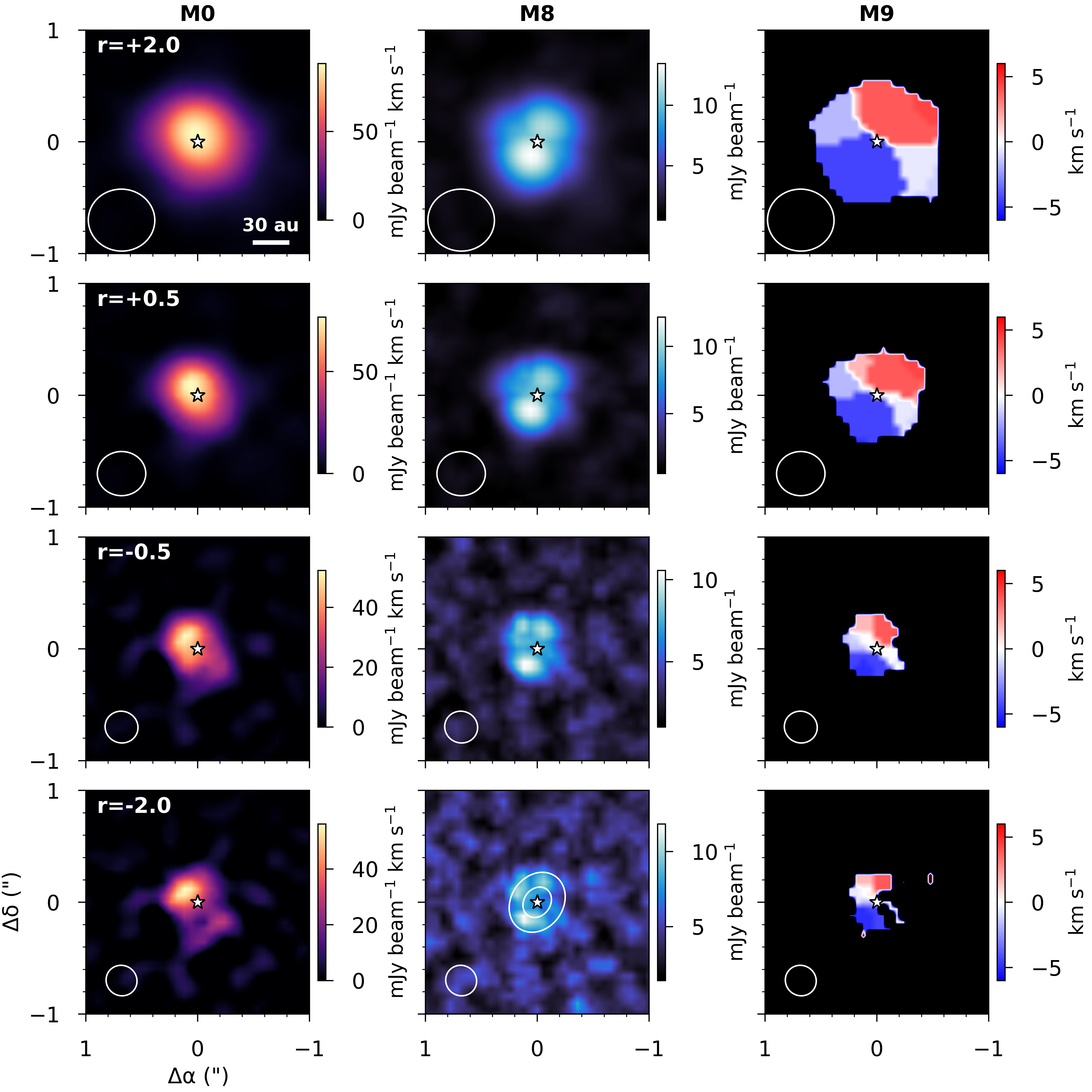}
    \caption{Moment maps of the stacked \ce{CH_3OH} lines imaged at different robust parameters [+2.0, +0.5,-0.5, -2.0]. From left to right the maps are the Keplerian masked integrated intensities (M0), peak intensity (M8) and peak velocity (M9) maps. In the center map the color bar starts at the 3~$\sigma$ level and the right maps have had a 5~$\sigma$ clip applied to the channel maps. The ellipses in the central panel of the bottom row highlight the ringed area used to calculate the column densities.}
    \label{fig:4}
\end{figure*}


\begin{table}
    \caption{Properties of the stacked \ce{CH_3OH} images}   
    \centering
    \begin{tabular}{c c c c c}
    \hline
    robust & Beam & rms$^{*}$ & Peak Intensity \\  
    & ($\farcs \times \farcs$ ($^{\circ}$)) & (mJy beam$^{-1}$) & (mJy beam$^{-1}$) \\ \hline \hline 
     +2.0    &  0.60$\times$0.55 (87) & 0.50  & 13.44 \\  
     +0.5    &  0.43$\times$0.39 (-90) & 0.55 & 12.18  \\ 
     -0.5    &  0.30$\times$0.28 (70) & 1.02  & 10.55  \\ 
     -2.0    &  0.28$\times$0.27 (56) & 1.41  & 11.84 \\    \hline 
    \end{tabular}
\tablecomments{$^{*}$per 0.55~km~s$^{-1}$ channels.}
    \label{stacked_image_properties}
\end{table}

\begin{figure}
    \centering
    \includegraphics[width=\linewidth]{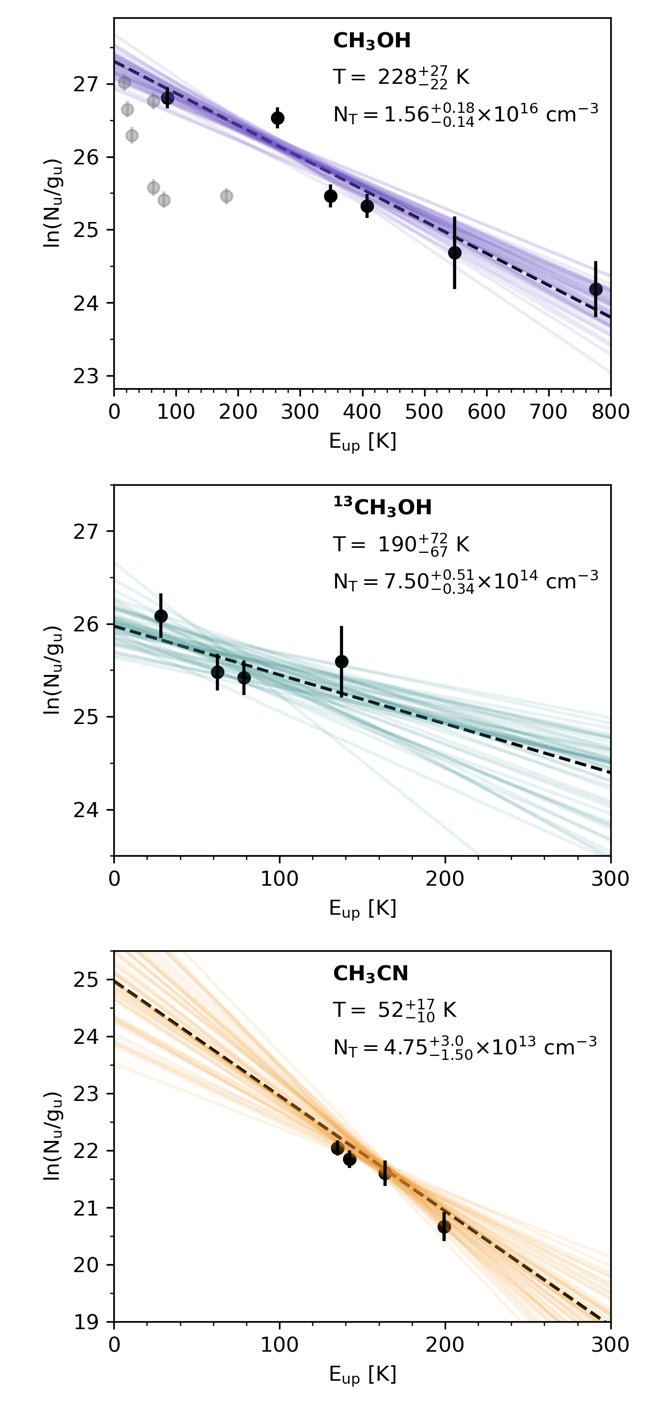}
    \caption{Disk averaged rotation diagrams for \ce{CH_3OH}, \ce{^{13}CH_3OH} and \ce{CH_3CN} in the HD~100453 disk. The gray points in the \ce{CH_3OH} plot show optically thicker transitions which are excluded from the linear fit. The dashed black lines show the best fit model and the colors show random draws from the corresponding posterior probability distributions.}
    \label{fig:rotation_diagram}
\end{figure}

\subsection{Rotational temperatures and column densities}

Due to the number of lines of \ce{CH_3OH}, \ce{^{13}CH_3OH} and \ce{CH_3CN} detected and the range in upper energy levels these transitions span we are able to calculate empirical rotational temperatures and column densities for each of these species (as listed in Table~\ref{table:2}). We iterate these calculations such that the thermal line width used to determine the optical depth is consistent with the recovered rotational temperature. 
The equations for the optical depth and thermal linewidth are fully described in \citet{2018ApJ...859..131L} and \citet{2020ApJ...890..142P}.
Within this framework we require an underlying emitting area of the gas which we assume is the same for all of the organic molecules. Given the sharp drop off of in the \ce{CH_3OH} line wings we set the inner radius of this area to be the inner edge of the mm-dust  and the outer radius to be half of the major axis for the beam for the $\mathrm{briggs~robust=-2.0}$ stacked \ce{CH_3OH} image (as shown in Figure~\ref{fig:4}). This results in a ring of gas from 16 to 30~au (note the size of the \ce{C^{18}O} gas disk is $\approx$60~au; \citealt{2024A&A...682A.149S}). Although there are hints at non-axisymmetric emission structures in these higher resolution images we do not have the sensitivity to determine this on a line-by-line basis and therefore adopt to a simple disk averaged column density approach. Our absolute column densities will therefore scale inversely with the chosen emitting area, as is typical, but we are most interested in the abundance ratios between different species and the rotational temperatures which are not effected by this assumption when in the optically thin regime. 

In Figure~\ref{fig:rotation_diagram} we show the resulting rotational diagrams for \ce{CH_3OH}, \ce{^{13}CH_3OH}, and \ce{CH_3CN}. For \ce{CH_3OH} there is a clear scatter in the diagram likely due to a range of line opacities probed by our data. To mitigate this, we only include lines with Einstein A coefficients lower than 10$^{-4}$~s$^{-1}$ and/or upper energy levels greater than 300~K in the linear fit. These are highlighted in Figure~\ref{fig:rotation_diagram} via the color of the scatter points where black are included and gray are excluded. This sub-set of lines all have optical depths $<$0.09 and the \ce{CH_3OH} column density is well constrained to $\approx$1.6$\times10^{16}$~cm$^{-2}$ with a rotational temperature of $\approx$230~K. The \ce{^{13}CH_3OH} fit retrieves a column density of $\approx$7.5$\times10^{14}$~cm$^{-2}$ and a somewhat lower temperature, but when considering the uncertainties it is consistent with that for \ce{CH_3OH}. 
We discuss the \ce{^{12}C}/\ce{^{13}C} ratio further in Section 4.3.
The results for \ce{CH_3CN} are notably different with a cooler rotational temperature of $\approx$50~K and a much lower column density of $\approx$4.0$\times10^{13}$~cm$^{-2}$. We attempted the same analysis for \ce{H_2CO} but the 4(1,3)-3(1,2) and 4(2,2)-3(2,1) transitions were found to be optically thick with $\tau \geqslant 1$. For the column density estimates of \ce{H_2CO} and \ce{CH_3OCHO}, \ce{CH_2DOH} and \ce{H_2^{13}CO}, as well as the upper-limits on \ce{HDO} and \ce{CH_3OCH_3} we calculate these at the fixed temperature of 228~K determined from the \ce{CH_3OH} as we expect these species to sublimate at approximate the same region of the disk. These results are listed in Table~\ref{table:2} alongside the relative percentage abundances of these species with respect to \ce{CH_3OH}. We find all of the species are present at less than the 5\% level. 


\begin{table*}[]
\caption{Derived column densities, abundance ratios and rotational temperatures for molecules the HD~100453 disk.}
\centering
\begin{tabular}{ccccc}
 \hline
Molecule            & $\mathrm{T_{rot}}$[K] & $\mathrm{N_{col}}$ [$\mathrm{cm^{-2}}$] & $\mathrm{N_{col}}$ @ 228~K$^{*}$~[$\mathrm{cm^{-2}}$]) 
& \% of \ce{CH_3OH} $\mathrm{N_{col}}$\\ \hline \hline
\ce{CH_3OH}         &  $228^{+27}_{22}$   &  1.56$^{+0.18}_{-0.14}\times10^{16}$     &      -  & -                     \\
\ce{^{13}CH_3OH}    &  $190^{+72}_{-67}$  & 7.50$^{+0.51}_{-0.34}\times10^{14}$     &   -   &   4.8$\pm$0.6                  \\
\ce{CH_2DOH}        &  -  &         -     &    6.67$^{+1.59}_{-1.58}\times10^{14}$   &   4.3$\pm$1.4                  \\
\ce{H_2CO}          &  - &  -             &     6.13$^{+0.55}_{-0.55}\times10^{14}$ &    3.9$\pm$0.5      \\
\ce{H_2^{13}CO}     & -   & -    &     3.55$^{+1.22}_{-1.22}\times10^{13}$        &   0.2$\pm$0.1  \\
\ce{CH_3CN}         &  $50^{+17}_{-10}$   &  4.74$^{+2.00}_{-1.10}\times10^{13}$    &    -   &   0.3$\pm$0.1                   \\
\ce{CH_3OCHO}       &  -   &  -    &      2.88$^{+0.33}_{-0.25}\times10^{14}$            &   1.9$\pm$0.2 \\
\ce{CH_3OCH3}       & -   & -    & $<3.70\times10^{14}$  &  $<$2.4 \\
\ce{HDO}       & -   & -    & $<6.25\times10^{14}$    &  $<$4.0\\ \hline
\end{tabular}
\label{table:2}
\tablecomments{$^{*}$ where 228~K is the rotational temperature derived from the \ce{CH_3OH}.}
\end{table*}

\section{Discussion} \label{sec:discussion} 

\begin{figure*}
    \centering
    \includegraphics[width=0.85\linewidth]{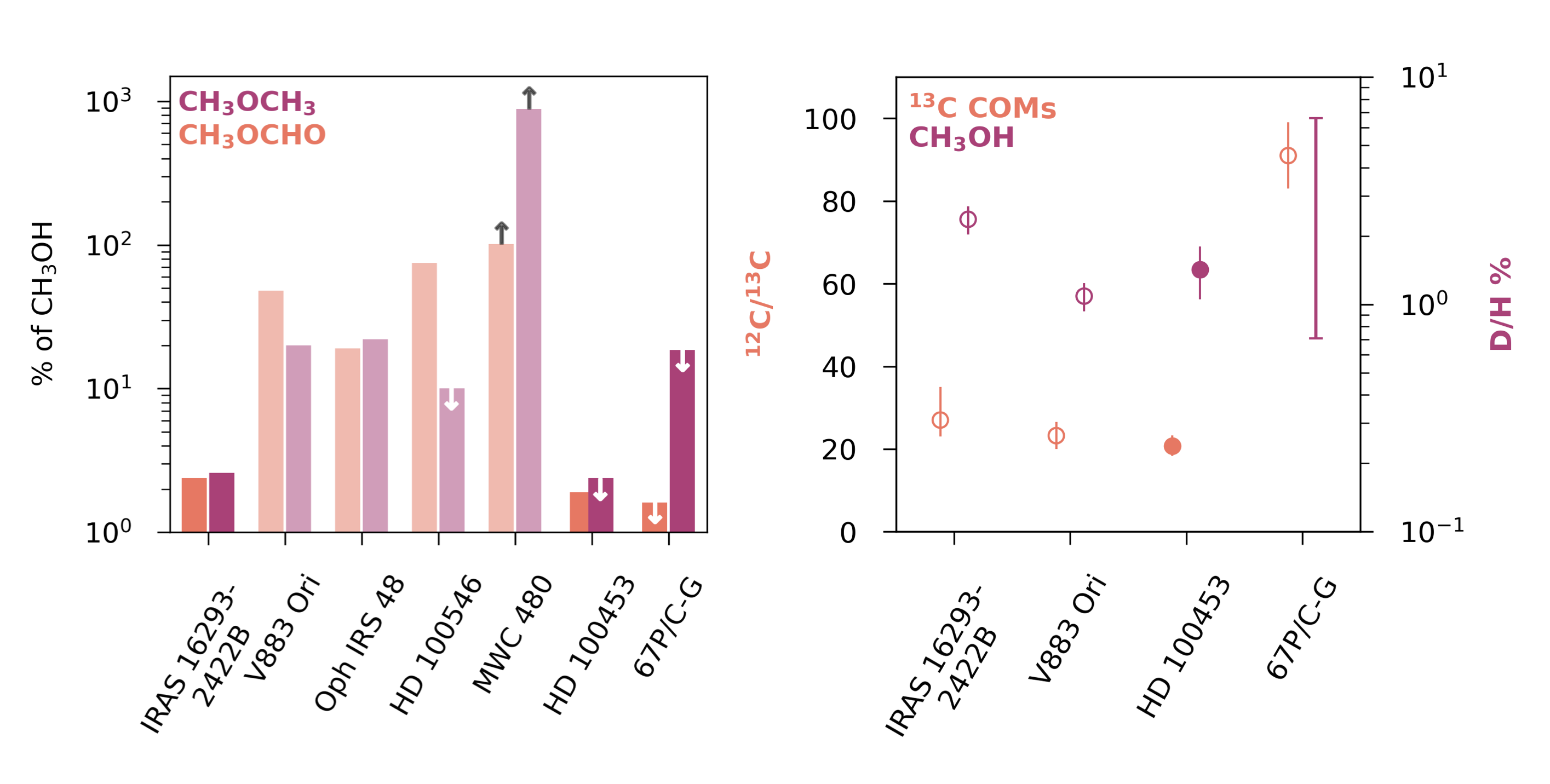}
    \caption{The abundances of various molecules in the HD~100453 disk compared to a selection of Class 0, I, and II objects as well as the solar system comet 67P/C-G. Left: The abundances of \ce{CH_3OCH_3} and \ce{CH_3OCHO} relative to \ce{CH_3OH}, where arrows denote upper and lower limits. Here, the Class II disks, aside from HD100453, are translucent to indicate the uncertainty in the underlying \ce{CH_3OH} column density. Right: The isotopic ratio of \ce{^{12}C}/\ce{^{13}C} for COMs (all \ce{CH_3OH} aside from IRAS~16293-2422B which is glycolaldehyde) and percentage D/H values for \ce{CH_3OH}. The filled circles show the results from this work, and the open circles are values taken from the literature. For both panels, the data for the other sources are taken from: \citet{2016A&A...595A.117J,2018A&A...620A.170J,2019MNRAS.489..594R, 2020MNRAS.498.5855A,2021MNRAS.500.4901D, 2024ApJ...974...83Y,2024AJ....167..164B, 2024AJ....167..165B,2025ApJS..276...49J}.}
    \label{fig:enter-label}
\end{figure*}

\subsection{The physical origin of COMs in the HD~100453 disk}

Our results show that the HD~100453 disk has a rich reservoir of complex organic molecules. The \ce{CH_3OH} line wings and rotational temperature corroborate that we are likely tracing a ring of gas originating from the thermal sublimation of \ce{CH_3OH}-rich ices at the inner dust cavity edge, very similar to the scenario first proposed for the HD~100546 disk \citep{2021NatAs...5..684B}. Looking at complementary tracers of the disk gas temperature, the optically thick \ce{^{12}CO} lines do reach brightness temperatures of $\approx$100~K in the vicinity of the dust cavity \citep{2023A&A...670A.154W}. Given these lines of evidence we infer that the \ce{CH_3OH} isotopologues and \ce{CH_3OCHO} detected in the HD~100453 system also originate from thermally sublimating ices close to the dust cavity wall. The lower rotational temperature of \ce{CH_3CN} relative to \ce{CH_3OH} may indicate a different physical origin for this species that will discussed further in Section 4.2. 

Notably, \citet{2023A&A...670A.154W} show a temperature asymmetry in the HD~100453 disk where the CO gas in the North East of the disk is warmer. Interestingly, this is the same region of the disk where we see the excess \ce{CH_3OH} when imaging the data at higher angular resolution. In addition, there is an over-brightness in the disk continuum emission at this same azimuth. These asymmetries could be linked to shadowing from the inner disk which is misaligned in both position and inclination angles \citep{2022A&A...658A.183B}. The locations of the disk shadows due to this misalignment are qualitatively rotated $\approx$90\degree~from the CO and \ce{CH_3OH} hot-spots which, supports this hypothesis, if the molecule excesses are tracing the warmer gas outwith the shadows. If the variation the thermal structure of the disk is significant enough this would lead to a cycle of desorption and adsorption of volatiles. Depending on the timescales that the gas reacts to this change, it may be possible to see time variable line emission from COMs as the orbital timescale at 0.315~au, the peak emitting radius of the inner disk, is only 0.14~years \citep{2022A&A...658A.183B}.

On larger scales, the HD~100453 disk is also known to have two spiral arms that are induced by an external binary companion \citep[e.g.;][]{2018ApJ...854..130W,2020MNRAS.491.1335R}. In gas emission, these have been traced by optically thick CO lines therefore, its hard to determine if this is a total gas density enhancement or a temperature enhancement. In these data, we do not see any clear signatures of spirals traced in molecular lines but given the spatial resolution of these data this is not surprising. 
Interestingly, the \ce{^{12}CO} J=3-2 spectra also shows the same apparent excess in emission close to the source velocity as the \ce{CH_3OH}. Therefore, instead of a small scale temperature asymmetry near the dust cavity edge, as discussed above, this spectral feature could be linked to the spiral arms.
The \ce{CH_3OH} and \ce{H_2CO} line strengths do mean that future high angular resolution ($\approx$0\farcs1) observations are feasible and this would enable a more detailed analysis into how we can use these molecules as density and temperature probes in this disk. 

\subsection{Placing the HD~100453 COMs reservoir in context}

The combination of high upper energy level $>$300~K and low Einstein A coefficients ($<$10$^{-4}$~s$^{-1}$) \ce{CH_3OH} lines detected in this study allow us to make a robust measurement of the \ce{CH_3OH} column density from optically thin lines. As a result, HD~100453 has at least an order of magnitude higher \ce{CH_3OH} column density when compared to other Class II disks, where reported values range from $\approx$\ce{10^{14}}-\ce{10^{15}} $\mathrm{cm^{-2}}$ \citep[][]{2023A&A...678A.146B,2024AJ....167..165B,2024arXiv241112418T,2025arXiv250204957E}. 
The rotational temperature derived for \ce{CH_3OH} in HD~100453 is warmer than the measurements for both the HD~100546 ($\approx$150~K) and Oph-IRS 48 ($\approx$125~K) disks \citep{2024arXiv241112418T,2025arXiv250204957E}.
The absolute values on the column densities do depend on assumptions regarding the emitting area but the line coverage of the data here shows that these weaker lines are detectable in Class II disks and therefore, the column densities from these other studies may be underestimates. These sources should be revisited with more optically thin observations since abundances extracted from optically thick and optically thin lines have different dependencies on assumed emission areas.

In HD~100453, we find that \ce{CH_3OCHO} is only present at the $\approx$2\% level of \ce{CH_3OH} which is at least a factor of 10 lower than that measured towards other Class II disks and at least a factor few lower than that seen towards younger out-bursting sources \citep{2024AJ....167..164B,2024AJ....167..165B,2024AJ....167...66Y,2024ApJ...974...83Y,2024ApJ...975..170C, 2025ApJS..276...49J}. 
\ce{CH_3OCH_3} is not detected in the HD~100453 disk, but the upper-limits relative to \ce{CH_3OCHO} are not constraining. 
The difference in the HD~100453 COMs abundances relative to \ce{CH_3OH} is shown in Figure~6 where we compare the \ce{CH_3OCHO} and \ce{CH_3OCH_3} column densities to those reported for three other Class II disks (HD~100546, Oph-IRS~48 and MWC~480), IRAS~16293-2422B, V883~Ori and comet 67P/C-G \citep[where column destinies are taken from][]{2016A&A...595A.117J,2018A&A...620A.170J,2019MNRAS.489..594R,2024AJ....167..164B, 2024AJ....167..165B,2025ApJS..276...49J}.
This percent level abundance is more consistent with the values seen in the warm gas phase towards younger systems \citep[e.g.,][]{2017ApJ...841..120B,2020A&A...639A..87V,2020A&A...635A..48M}. This result may support the intact inheritance of COMs in disks rather than the in-situ processing of inherited \ce{CH_3OH} into other COMs as proposed for the other Class II disks \citep[e.g.;][]{2024AJ....167..164B,2024AJ....167..165B}. Its not clear yet what the intrinsic variation in COMs reservoirs across Class II disks is and, if the high abundances seen in other works my be due to an under-appreciation of optical depth effects impacting the derived \ce{CH_3OH} column densities. 

After TW~Hya, HD~100453 is the second Class II disk so far where both \ce{CH_3OH} and \ce{CH_3CN} have been detected. Looking at the relative abundance of \ce{CH_3CN} seen towards protostars, our \ce{CH_3CN}/\ce{CH_3OH} abundance of $\approx$0.3\% is within the ranges seen in this warmer environment \citep[e.g.,][]{2017ApJ...841..120B,2021A&A...650A.150N} and approximately an order of magnitude lower than that of 67P/C-G \citep{2019MNRAS.489..594R}.  When comparing HD~100453 to the transition disks HD~100546 and Oph-IRS~48, where \ce{CH_3CN} has gone undetected, the upper-limits relative to \ce{CH_3OH} show that deeper data are needed to detected this potential reservoir of nitrogen-rich COMs in these other disks \citep{2024AJ....167..164B,2024AJ....167..165B}. Interestingly, the rotational temperature for \ce{CH_3CN} in HD~100453 is significantly cooler than that for \ce{CH_3OH}, bringing into question if this is indeed an ice sublimation product, given that these two species have similar binding energies \citep{2022ESC.....6..597M}. Instead these molecules may be tracing different radial and/or vertical regions of the disk. 

The presence of COMs in Class II disk have also been used as tracers of the gas-phase C/O, in particular, the molecules \ce{CH_3OH} and \ce{CH_3CN} have been shown to trace environments of C/O$\approx$0.5 and C/O$>$1 respectively \citep{2023A&A...678A.146B,2023NatAs...7...49C}.
The lack of \ce{C_2H} in the HD~100453 system \citep{2022MNRAS.510.1148S} indicates that the disk gas has a C/O$\leq$1, i.e., there is no relative excess of gas-phase carbon as seen in other systems. The detection of \ce{CH_3CN} therefore may be somewhat at odds with the scenario presented in \citep{2023NatAs...7...49C}. We interpret the cooler \ce{CH_3CN} gas as originating from a different physical region of the disk to the \ce{CH_3OH}. Therefore, we could be seeing a radial change in C/O where the drop in rotational temperature of the \ce{CH_3CN} compared to the \ce{CH_3OH} indicates that this species is tracing cooler gas beyond the water snowline where C/O$\approx$0.8-1.0.


\subsection{Exploring the \ce{^{13}C} ratio in disk organics}

With the detection of \ce{^{13}CH_3OH} we now have a window into the \ce{^{12}C}/\ce{^{13}C} ratio in the disk's large organic reservoir. In Class II disks the most complete information we have on the variation of \ce{^{12}C}/\ce{^{13}C} across different molecular families is from observations of the TW-Hya protoplanetary disk as collated by \citet{2024ApJ...965..147B}. In TW-Hya there are two distinct isotopic reservoirs for carbon where the CO gas is enhanced in \ce{^{13}C} within 110~au whereas within the same radii the \ce{C_2H}, CN and HCN molecules are consistent with the local ISM abundances or somewhat depleted in \ce{^{13}C} \citep{2017NatAs...1E.130Z,2019A&A...632L..12H,2022ApJ...937L..14Y,2024ApJ...966...63Y,2024ApJ...965..147B}. This fractionation of carbon is proposed to happen in the gas-phase and depending on the height in the disk is driven by isotope-selective photo-dissociation or cosmic-ray
ionization closer to the mid-plane where the low-temperature carbon isotopic fractionation reaction(s) are active \citep{1980ApJ...242..424S, 2009ApJ...693.1360W,2024ApJ...965..147B}. 
There is also a C/O dependence where the gas-phase isotope exchange reactions are more efficient in regimes there C/O$>$1 \citep{1984ApJ...277..581L,2022ApJ...932..126Y,2024ApJ...969...41L}.
For the larger organics that are chemically derived from CO, such as \ce{H_2CO} and \ce{CH_3OH}, which can form efficiently on grain surfaces, the potential in-situ carbon isotopic fractionation in disks has yet to be explored in detail either from an observational or a modeling perspective. \ce{H_{2}^{13}CO} has been detected in the Oph-IRS~48 and HD~100546 disks \citep{2024AJ....167..164B,2024AJ....167..165B}, and for HD~100546 in particular, the \ce{H_{2}^{13}CO} is detected in the outer disk ring where the \ce{H_2CO} is optically thin and cold $\approx$30~K \citep{2025arXiv250204957E}. 
From the HD~100546 data presented in \citet{2024AJ....167..164B} there is the indication of a $\approx$2$\times$ enhancement in \ce{^{13}C} in \ce{H_2CO} relative to the ISM, but higher signal-to-noise data are needed to confirm. 

For the HD~100453 disk we can measure the \ce{^{12}C}/\ce{^{13}C} ratio in CO, \ce{H_2CO}, and \ce{CH_3OH}. Unfortunately, the isotopic ratios in CO and \ce{H_2CO} are difficult to constrain here due to the high optical depth of \ce{C^{18}O} \citep[e.g.;][]{2024A&A...682A.149S} and due to the tentative nature of the \ce{H_2^{13}CO} detection. From the disk integrated column densities we find a lower-limit \ce{^{12}C}/\ce{^{13}C} in \ce{H_2CO} of $>$17. For \ce{CH_3OH} we find a \ce{^{12}C}/\ce{^{13}C} ratio of 21$\pm$3 which is a $\approx$3$\times$ enhancement in \ce{^{13}C} relative to the local ISM value of 69 \citep{1999RPPh...62..143W}. As shown in Figure~6, this level of enhancement in \ce{^{13}C} has also been seen in the \ce{CH_3OH} and other COMs from the out-bursting Class I source V883~Ori \citep{2025ApJS..276...49J, 2024AJ....167...66Y}, possibly in the \ce{CH_3OH} in the Class 0 HH~212 disk \citep{2019ApJ...876...63L}, and in the glycolaldehyde detected in the Class O IRAS 16293-2422~B \citep{2016A&A...595A.117J}. Interestingly, in comparison, the \ce{^{13}C} in comet 67P/C-G is consistent with the solar value of 91 which is somewhat depleted relative to the local ISM \citep{2020MNRAS.498.5855A}. 
Such a clear signature of isotopic fractionation in the ices in molecular clouds and protostellar envelopes have so far eluded detection in CO and \ce{CO_2} ices \citep{2023NatAs...7..431M, 2024A&A...692A.163B}. This indicates that that the enhancement of \ce{^{13}C} in CO and other CO derived organics may take place within the disk environment rather than being inherited from earlier times and/or there are different isotopic reservoirs in the simple molecules versus the (complex) organics.  
The enrichment of \ce{^{13}C} in organics may provide an answer to the overabundant \ce{^{13}CO} traced in giant exoplanet atmospheres \citep{2021Natur.595..370Z}.


\subsection{Are we tracing inherited ices?}

The level of inheritance for disk organics can be evaluated based on the overall organic composition, and the isotopic ratios among key organics. In addition to \ce{CH_3OH} and \ce{^{13}CH_3OH} we also identify \ce{CH_2DOH} for the first time in a Class II disk. From these data we find a D/H value in \ce{CH_3OH} of 1-2\% which, as shown in Figure~6, is consistent with the deuterium enhancement in \ce{CH_3OH} observed towards pre-stellar cores and protostars \citep[e.g.;][]{2021MNRAS.501..347A,2020A&A...635A..48M,2022A&A...667A.136V,2023A&A...669L...6L} as well as the young disk V883~Ori, the HH~212 disk, and solar system comets \citep{2019ApJ...876...63L,2021MNRAS.500.4901D,2025ApJS..276...49J}. 
Given the low formation efficiency of \ce{CH_3OH} in warm Herbig disks \citep{2021NatAs...5..684B} the in-situ deuteration of \ce{CH_3OH} is unlikely but should be tested with future chemical modelling efforts. 
That being said, this result adds to growing evidence that Class II disks at least partially inherit icy material from the earlier stages of star formation.

Given the rotational temperature of the \ce{CH_3OH} we are likely tracing gas within the \ce{H_2O} snowline and thus \ce{H_2O} should be observable in the gas-phase too. 
We report an upper limit on the heavy water isotopologue HDO of $<$4\% that of \ce{CH_3OH}. Comparing this to the detections in protostars, outbursting sources and comets \citep[as collated in][]{2024ApJ...975..170C} these data are not sensitive enough to trace this avenue of ice inheritance. Future data to target more favorable \ce{H_2O} (isotopologue) lines as predicted by \citet{2019ApJ...875...96N} and seen in other systems \citep{2023Natur.615..227T,2024NatAs...8..587F} are necessary to make this much needed connection. 

\section{Conclusion} \label{sec:conclusion} 

In this work, we have presented the analysis of line rich ALMA observations towards the HD~100453 protoplanetary disk. Our main conclusions are as follows:

\begin{itemize}
    \item The abundant \ce{CH_3OH} detected is consistent with a ring of thermally sublimating ices located at the inner edge of the dust ring at $\approx$16~au. 
    This \ce{CH_3OH} emission is likely originating from within the \ce{H_2O} snowline and is tracing gas with a C/O$<$1. Additionally, there is evidence for an over-brightness along the disk minor axis that may be linked to shadowing due to a known misaligned inner disk on sub-au scales.
    
    \item For the first time in a Class II disk we detect \ce{^{13}CH_3OH} and infer from this an enhanced \ce{^{13}C} abundance relative the local ISM. This $\approx3\times$ increase is consistent with the high \ce{^{13}C} content of COMs in the V883 Ori disk, the CO gas in the TW Hya disk, and the CO measurements of gas-giant exoplanet atmospheres. 
    
    \item With the tentative detection of \ce{CH_2DOH} we can also make the first measurement of the D/H ratio in Class II disk organics. We find a D/H value of $\approx$1-2\% which is consistent with what is seen for \ce{CH_3OH} towards both low mass pre-stellar cores protostars, young Class I disks and, solar system comets adding further evidence to support the presence of inherited interstellar ices in disks. 

    \item The detection of \ce{CH_3CN} makes HD~100453 only the second Class II disk where both \ce{CH_3CN} and \ce{CH_3OH} have been detected. Interestingly, the \ce{CH_3CN} is emitting from a much cooler gas reservoir than the \ce{CH_3OH}, consistent with what has been measured for other Class II disks where the C/O$>$1 rather than the oxygen-rich gas present in this system. 
    
     \item  The larger COM \ce{CH_3OCHO} is detected but the relative abundance with respect to \ce{CH_3OH} is an order of magnitude lower than that reported for other Class II disks. In comparison, this lower value is more consistent with younger proto-stellar systems and may be due to the detection of weaker and higher energy level transitions of \ce{CH_3OH} which lead to an overall higher reported \ce{CH_3OH} column density.  
    
    \item Finally, we report a non-detection of the heavy water isotopologue HDO and find that the derived upper-limit is not constraining when compared to the abundances measured in other interstellar environments. Future data targeting more favorable transitions are needed to elucidate the nature of the cool water reservoir in disks as probed by ALMA. 
    
\end{itemize}
HD~100453 is now the fifth disk around a Herbig Ae (or young F-type) star where thermally sublimating COMs have been detected \citep{2021NatAs...5..684B, 2021A&A...651L...5V, 2023A&A...678A.146B,2024ApJ...974...83Y}. These disks clearly give us access to a new complex organic and isotopic reservoir that has so far been inaccessible in cooler and lower mass T-Tauri disks. Therefore, ALMA observations of these systems are crucial in unraveling the history and inheritance of organics through the star, disk and planet formation process.

\begin{acknowledgements}

This paper makes use of the following ALMA data: 2017.1.01424.S, 2023.1.00252.S. ALMA is a partnership of ESO (representing its member states), NSF (USA) and NINS (Japan), together with NRC (Canada), MOST and ASIAA (Taiwan), and KASI (Republic of Korea), in cooperation with the Republic of Chile. The Joint ALMA Observatory is operated by ESO, AUI/NRAO and NAOJ. 
This work has used the following additional software packages that have not been referred to in the main text: Astropy, IPython, Jupyter, Matplotlib and NumPy \citep{Astropy,IPython,Jupyter,Matplotlib,NumPy}.
A.S.B. is supported by a Clay Postdoctoral Fellowship from the Smithsonian Astrophysical Observatory. 
M.L. is funded by the European Union (ERC, UNVEIL, 101076613). Views and opinions expressed are however those of the author(s) only and do not necessarily reflect those of the European Union or the European Research Council. Neither the European Union nor the granting authority can be held responsible for them.
Support for C.J.L. was provided by NASA through the NASA Hubble Fellowship grant No. HST-HF2- 51535.001-A awarded by the Space Telescope Science Institute, which is operated by the Association of Universities for Research in Astronomy, Inc., for NASA, under contract NAS5-26555.
S.N. is grateful for support from Grants-in-Aid for JSPS (Japan Society for the Promotion of Science) Fellows Grant Number JP23KJ0329, MEXT/JSPS Grants-in-Aid for Scientific Research (KAKENHI) Grant Numbers JP23K13155 and JP24K00674, and Start-up Research Grant as one of The University of Tokyo Excellent Young Researcher 2024.
C.W.~acknowledges financial support from the Science and Technology Facilities Council and UK Research and Innovation (grant numbers ST/X001016/1 and MR/T040726/1).

\end{acknowledgements}

\newpage
\bibliography{sample631}{}

\newcommand{\noopsort}[1]{} \newcommand{\printfirst}[2]{#1}
  \newcommand{\singleletter}[1]{#1} \newcommand{\switchargs}[2]{#2#1}
\begin{thebibliography}{}
\expandafter\ifx\csname natexlab\endcsname\relax\def\natexlab#1{#1}\fi
\providecommand{\url}[1]{\href{#1}{#1}}
\providecommand{\dodoi}[1]{doi:~\href{http://doi.org/#1}{\nolinkurl{#1}}}
\providecommand{\doeprint}[1]{\href{http://ascl.net/#1}{\nolinkurl{http://ascl.net/#1}}}
\providecommand{\doarXiv}[1]{\href{https://arxiv.org/abs/#1}{\nolinkurl{https://arxiv.org/abs/#1}}}

\bibitem[{{Allamandola} {et~al.}(1988){Allamandola}, {Sandford}, \&
  {Valero}}]{1988Icar...76..225A}
{Allamandola}, L.~J., {Sandford}, S.~A., \& {Valero}, G.~J. 1988, \icarus, 76,
  225, \dodoi{10.1016/0019-1035(88)90070-X}

\bibitem[{{Altwegg} {et~al.}(2020){Altwegg}, {Balsiger}, {Combi}, {De Keyser},
  {Drozdovskaya}, {Fuselier}, {Gombosi}, {H{\"a}nni}, {Rubin}, {Schuhmann},
  {Schroeder}, \& {Wampfler}}]{2020MNRAS.498.5855A}
{Altwegg}, K., {Balsiger}, H., {Combi}, M., {et~al.} 2020, \mnras, 498, 5855,
  \dodoi{10.1093/mnras/staa2701}

\bibitem[{{Ambrose} {et~al.}(2021){Ambrose}, {Shirley}, \&
  {Scibelli}}]{2021MNRAS.501..347A}
{Ambrose}, H.~E., {Shirley}, Y.~L., \& {Scibelli}, S. 2021, \mnras, 501, 347,
  \dodoi{10.1093/mnras/staa3649}

\bibitem[{{Astropy Collaboration} {et~al.}(2022){Astropy Collaboration},
  {Price-Whelan}, {Lim}, {Earl}, {Starkman}, {Bradley}, {Shupe}, {Patil},
  {Corrales}, {Brasseur}, {N{\"o}the}, {Donath}, {Tollerud}, {Morris},
  {Ginsburg}, {Vaher}, {Weaver}, {Tocknell}, {Jamieson}, {van Kerkwijk},
  {Robitaille}, {Merry}, {Bachetti}, {G{\"u}nther}, {Aldcroft},
  {Alvarado-Montes}, {Archibald}, {B{\'o}di}, {Bapat}, {Barentsen},
  {Baz{\'a}n}, {Biswas}, {Boquien}, {Burke}, {Cara}, {Cara}, {Conroy},
  {Conseil}, {Craig}, {Cross}, {Cruz}, {D'Eugenio}, {Dencheva}, {Devillepoix},
  {Dietrich}, {Eigenbrot}, {Erben}, {Ferreira}, {Foreman-Mackey}, {Fox},
  {Freij}, {Garg}, {Geda}, {Glattly}, {Gondhalekar}, {Gordon}, {Grant},
  {Greenfield}, {Groener}, {Guest}, {Gurovich}, {Handberg}, {Hart},
  {Hatfield-Dodds}, {Homeier}, {Hosseinzadeh}, {Jenness}, {Jones}, {Joseph},
  {Kalmbach}, {Karamehmetoglu}, {Ka{\l}uszy{\'n}ski}, {Kelley}, {Kern},
  {Kerzendorf}, {Koch}, {Kulumani}, {Lee}, {Ly}, {Ma}, {MacBride}, {Maljaars},
  {Muna}, {Murphy}, {Norman}, {O'Steen}, {Oman}, {Pacifici}, {Pascual},
  {Pascual-Granado}, {Patil}, {Perren}, {Pickering}, {Rastogi}, {Roulston},
  {Ryan}, {Rykoff}, {Sabater}, {Sakurikar}, {Salgado}, {Sanghi}, {Saunders},
  {Savchenko}, {Schwardt}, {Seifert-Eckert}, {Shih}, {Jain}, {Shukla}, {Sick},
  {Simpson}, {Singanamalla}, {Singer}, {Singhal}, {Sinha}, {Sip{\H{o}}cz},
  {Spitler}, {Stansby}, {Streicher}, {{\v{S}}umak}, {Swinbank}, {Taranu},
  {Tewary}, {Tremblay}, {de Val-Borro}, {Van Kooten}, {Vasovi{\'c}}, {Verma},
  {de Miranda Cardoso}, {Williams}, {Wilson}, {Winkel}, {Wood-Vasey}, {Xue},
  {Yoachim}, {Zhang}, {Zonca}, \& {Astropy Project Contributors}}]{Astropy}
{Astropy Collaboration}, {Price-Whelan}, A.~M., {Lim}, P.~L., {et~al.} 2022,
  \apj, 935, 167, \dodoi{10.3847/1538-4357/ac7c74}

\bibitem[{{Benisty} {et~al.}(2017){Benisty}, {Stolker}, {Pohl}, {de Boer},
  {Lesur}, {Dominik}, {Dullemond}, {Langlois}, {Min}, {Wagner}, {Henning},
  {Juhasz}, {Pinilla}, {Facchini}, {Apai}, {van Boekel}, {Garufi}, {Ginski},
  {M{\'e}nard}, {Pinte}, {Quanz}, {Zurlo}, {Boccaletti}, {Bonnefoy}, {Beuzit},
  {Chauvin}, {Cudel}, {Desidera}, {Feldt}, {Fontanive}, {Gratton}, {Kasper},
  {Lagrange}, {LeCoroller}, {Mouillet}, {Mesa}, {Sissa}, {Vigan}, {Antichi},
  {Buey}, {Fusco}, {Gisler}, {Llored}, {Magnard}, {Moeller-Nilsson}, {Pragt},
  {Roelfsema}, {Sauvage}, \& {Wildi}}]{2017A&A...597A..42B}
{Benisty}, M., {Stolker}, T., {Pohl}, A., {et~al.} 2017, \aap, 597, A42,
  \dodoi{10.1051/0004-6361/201629798}

\bibitem[{{Bergin} {et~al.}(2024){Bergin}, {Bosman}, {Teague}, {Calahan},
  {Willacy}, {Cleeves}, {Schwarz}, {Zhang}, \&
  {Bruderer}}]{2024ApJ...965..147B}
{Bergin}, E.~A., {Bosman}, A., {Teague}, R., {et~al.} 2024, \apj, 965, 147,
  \dodoi{10.3847/1538-4357/ad3443}

\bibitem[{{Bergner} {et~al.}(2018){Bergner}, {Guzm{\'a}n}, {{\"O}berg},
  {Loomis}, \& {Pegues}}]{2018ApJ...857...69B}
{Bergner}, J.~B., {Guzm{\'a}n}, V.~G., {{\"O}berg}, K.~I., {Loomis}, R.~A., \&
  {Pegues}, J. 2018, \apj, 857, 69, \dodoi{10.3847/1538-4357/aab664}

\bibitem[{{Bergner} {et~al.}(2017){Bergner}, {{\"O}berg}, {Garrod}, \&
  {Graninger}}]{2017ApJ...841..120B}
{Bergner}, J.~B., {{\"O}berg}, K.~I., {Garrod}, R.~T., \& {Graninger}, D.~M.
  2017, \apj, 841, 120, \dodoi{10.3847/1538-4357/aa72f6}

\bibitem[{{Bertin} {et~al.}(2016){Bertin}, {Romanzin}, {Doronin}, {Philippe},
  {Jeseck}, {Ligterink}, {Linnartz}, {Michaut}, \&
  {Fillion}}]{2016ApJ...817L..12B}
{Bertin}, M., {Romanzin}, C., {Doronin}, M., {et~al.} 2016, \apjl, 817, L12,
  \dodoi{10.3847/2041-8205/817/2/L12}

\bibitem[{{Bohn} {et~al.}(2022){Bohn}, {Benisty}, {Perraut}, {van der Marel},
  {W{\"o}lfer}, {van Dishoeck}, {Facchini}, {Manara}, {Teague}, {Francis},
  {Berger}, {Garcia-Lopez}, {Ginski}, {Henning}, {Kenworthy}, {Kraus},
  {M{\'e}nard}, {M{\'e}rand}, \& {P{\'e}rez}}]{2022A&A...658A.183B}
{Bohn}, A.~J., {Benisty}, M., {Perraut}, K., {et~al.} 2022, \aap, 658, A183,
  \dodoi{10.1051/0004-6361/202142070}

\bibitem[{{Booth} {et~al.}(2023){Booth}, {Law}, {Temmink}, {Leemker}, \&
  {Mac{\'\i}as}}]{2023A&A...678A.146B}
{Booth}, A.~S., {Law}, C.~J., {Temmink}, M., {Leemker}, M., \& {Mac{\'\i}as},
  E. 2023, \aap, 678, A146, \dodoi{10.1051/0004-6361/202346974}

\bibitem[{{Booth} {et~al.}(2021){Booth}, {Walsh}, {Terwisscha van Scheltinga},
  {van Dishoeck}, {Ilee}, {Hogerheijde}, {Kama}, \&
  {Nomura}}]{2021NatAs...5..684B}
{Booth}, A.~S., {Walsh}, C., {Terwisscha van Scheltinga}, J., {et~al.} 2021,
  Nature Astronomy, 5, 684, \dodoi{10.1038/s41550-021-01352-w}

\bibitem[{{Booth} {et~al.}(2024{\natexlab{a}}){Booth}, {Leemker}, {van
  Dishoeck}, {Evans}, {Ilee}, {Kama}, {Keyte}, {Law}, {van der Marel},
  {Nomura}, {Notsu}, {{\"O}berg}, {Temmink}, \& {Walsh}}]{2024AJ....167..164B}
{Booth}, A.~S., {Leemker}, M., {van Dishoeck}, E.~F., {et~al.}
  2024{\natexlab{a}}, \aj, 167, 164, \dodoi{10.3847/1538-3881/ad2700}

\bibitem[{{Booth} {et~al.}(2024{\natexlab{b}}){Booth}, {Temmink}, {van
  Dishoeck}, {Evans}, {Ilee}, {Kama}, {Keyte}, {Law}, {Leemker}, {van der
  Marel}, {Nomura}, {Notsu}, {{\"O}berg}, \& {Walsh}}]{2024AJ....167..165B}
{Booth}, A.~S., {Temmink}, M., {van Dishoeck}, E.~F., {et~al.}
  2024{\natexlab{b}}, \aj, 167, 165, \dodoi{10.3847/1538-3881/ad26ff}

\bibitem[{{Brunken} {et~al.}(2022){Brunken}, {Booth}, {Leemker}, {Nazari}, {van
  der Marel}, \& {van Dishoeck}}]{2022A&A...659A..29B}
{Brunken}, N. G.~C., {Booth}, A.~S., {Leemker}, M., {et~al.} 2022, \aap, 659,
  A29, \dodoi{10.1051/0004-6361/202142981}

\bibitem[{{Brunken} {et~al.}(2024){Brunken}, {van Dishoeck}, {Slavicinska}, {le
  Gouellec}, {Rocha}, {Francis}, {Tychoniec}, {van Gelder}, {Navarro},
  {Boogert}, {Kavanagh}, {Nazari}, {Greene}, {Ressler}, \&
  {Majumdar}}]{2024A&A...692A.163B}
{Brunken}, N.~G.~C., {van Dishoeck}, E.~F., {Slavicinska}, K., {et~al.} 2024,
  \aap, 692, A163, \dodoi{10.1051/0004-6361/202451794}

\bibitem[{{Calahan} {et~al.}(2024){Calahan}, {Bergin}, {van't Hoff}, {Booth},
  {{\"O}berg}, {Zhang}, {Calvet}, \& {Hartmann}}]{2024ApJ...975..170C}
{Calahan}, J.~K., {Bergin}, E.~A., {van't Hoff}, M., {et~al.} 2024, \apj, 975,
  170, \dodoi{10.3847/1538-4357/ad78d1}

\bibitem[{{Calahan} {et~al.}(2023){Calahan}, {Bergin}, {Bosman}, {Rich},
  {Andrews}, {Bergner}, {Cleeves}, {Guzm{\'a}n}, {Huang}, {Ilee}, {Law}, {Le
  Gal}, {{\"O}berg}, {Teague}, {Walsh}, {Wilner}, \&
  {Zhang}}]{2023NatAs...7...49C}
{Calahan}, J.~K., {Bergin}, E.~A., {Bosman}, A.~D., {et~al.} 2023, Nature
  Astronomy, 7, 49, \dodoi{10.1038/s41550-022-01831-8}

\bibitem[{{Carney} {et~al.}(2019){Carney}, {Hogerheijde}, {Guzm{\'a}n},
  {Walsh}, {{\"O}berg}, {Fayolle}, {Cleeves}, {Carpenter}, \&
  {Qi}}]{2019A&A...623A.124C}
{Carney}, M.~T., {Hogerheijde}, M.~R., {Guzm{\'a}n}, V.~V., {et~al.} 2019,
  \aap, 623, A124, \dodoi{10.1051/0004-6361/201834353}

\bibitem[{{Collins} {et~al.}(2009){Collins}, {Grady}, {Hamaguchi},
  {Wisniewski}, {Brittain}, {Sitko}, {Carpenter}, {Williams}, {Mathews},
  {Williger}, {van Boekel}, {Carmona}, {Henning}, {van den Ancker}, {Meeus},
  {Chen}, {Petre}, \& {Woodgate}}]{2009ApJ...697..557C}
{Collins}, K.~A., {Grady}, C.~A., {Hamaguchi}, K., {et~al.} 2009, \apj, 697,
  557, \dodoi{10.1088/0004-637X/697/1/557}

\bibitem[{{Drozdovskaya} {et~al.}(2014){Drozdovskaya}, {Walsh}, {Visser},
  {Harsono}, \& {van Dishoeck}}]{2014MNRAS.445..913D}
{Drozdovskaya}, M.~N., {Walsh}, C., {Visser}, R., {Harsono}, D., \& {van
  Dishoeck}, E.~F. 2014, \mnras, 445, 913, \dodoi{10.1093/mnras/stu1789}

\bibitem[{{Drozdovskaya} {et~al.}(2021){Drozdovskaya}, {Schroeder I}, {Rubin},
  {Altwegg}, {van Dishoeck}, {Kulterer}, {De Keyser}, {Fuselier}, \&
  {Combi}}]{2021MNRAS.500.4901D}
{Drozdovskaya}, M.~N., {Schroeder I}, I. R.~H.~G., {Rubin}, M., {et~al.} 2021,
  \mnras, 500, 4901, \dodoi{10.1093/mnras/staa3387}

\bibitem[{{Evans} {et~al.}(2025){Evans}, {Booth}, {Walsh}, {Ilee}, {Keyte},
  {Law}, {Leemker}, {Notsu}, {{\"O}berg}, {Temmink}, \& {van der
  Marel}}]{2025arXiv250204957E}
{Evans}, L., {Booth}, A.~S., {Walsh}, C., {et~al.} 2025, arXiv e-prints,
  arXiv:2502.04957, \dodoi{10.48550/arXiv.2502.04957}

\bibitem[{{Facchini} {et~al.}(2024){Facchini}, {Testi}, {Humphreys}, {Vander
  Donckt}, {Isella}, {Wrzosek}, {Baudry}, {Gray}, {Richards}, \&
  {Vlemmings}}]{2024NatAs...8..587F}
{Facchini}, S., {Testi}, L., {Humphreys}, E., {et~al.} 2024, Nature Astronomy,
  8, 587, \dodoi{10.1038/s41550-024-02207-w}

\bibitem[{{Favre} {et~al.}(2018){Favre}, {Fedele}, {Semenov}, {Parfenov},
  {Codella}, {Ceccarelli}, {Bergin}, {Chapillon}, {Testi}, {Hersant},
  {Lefloch}, {Fontani}, {Blake}, {Cleeves}, {Qi}, {Schwarz}, \&
  {Taquet}}]{2018ApJ...862L...2F}
{Favre}, C., {Fedele}, D., {Semenov}, D., {et~al.} 2018, \apjl, 862, L2,
  \dodoi{10.3847/2041-8213/aad046}

\bibitem[{{Foreman-Mackey} {et~al.}(2013){Foreman-Mackey}, {Hogg}, {Lang}, \&
  {Goodman}}]{2013PASP..125..306F}
{Foreman-Mackey}, D., {Hogg}, D.~W., {Lang}, D., \& {Goodman}, J. 2013, \pasp,
  125, 306, \dodoi{10.1086/670067}

\bibitem[{{Gerakines} {et~al.}(1996){Gerakines}, {Schutte}, \&
  {Ehrenfreund}}]{1996A&A...312..289G}
{Gerakines}, P.~A., {Schutte}, W.~A., \& {Ehrenfreund}, P. 1996, \aap, 312, 289

\bibitem[{{Guzm{\'a}n-D{\'\i}az} {et~al.}(2021){Guzm{\'a}n-D{\'\i}az},
  {Mendigut{\'\i}a}, {Montesinos}, {Oudmaijer}, {Vioque}, {Rodrigo}, {Solano},
  {Meeus}, \& {Marcos-Arenal}}]{2021A&A...650A.182G}
{Guzm{\'a}n-D{\'\i}az}, J., {Mendigut{\'\i}a}, I., {Montesinos}, B., {et~al.}
  2021, \aap, 650, A182, \dodoi{10.1051/0004-6361/202039519}

\bibitem[{Harris {et~al.}(2020)Harris, Millman, van~der Walt, Gommers,
  Virtanen, Cournapeau, Wieser, Taylor, Berg, Smith, Kern, Picus, Hoyer, van
  Kerkwijk, Brett, Haldane, del R{\'{i}}o, Wiebe, Peterson,
  G{\'{e}}rard-Marchant, Sheppard, Reddy, Weckesser, Abbasi, Gohlke, \&
  Oliphant}]{NumPy}
Harris, C.~R., Millman, K.~J., van~der Walt, S.~J., {et~al.} 2020, Nature, 585,
  357, \dodoi{10.1038/s41586-020-2649-2}

\bibitem[{{Herbst} \& {van Dishoeck}(2009)}]{2009ARA&A..47..427H}
{Herbst}, E., \& {van Dishoeck}, E.~F. 2009, \araa, 47, 427,
  \dodoi{10.1146/annurev-astro-082708-101654}

\bibitem[{{Hily-Blant} {et~al.}(2019){Hily-Blant}, {Magalhaes de Souza},
  {Kastner}, \& {Forveille}}]{2019A&A...632L..12H}
{Hily-Blant}, P., {Magalhaes de Souza}, V., {Kastner}, J., \& {Forveille}, T.
  2019, \aap, 632, L12, \dodoi{10.1051/0004-6361/201936750}

\bibitem[{Hunter(2007)}]{Matplotlib}
Hunter, J.~D. 2007, Computing in Science \& Engineering, 9, 90,
  \dodoi{10.1109/MCSE.2007.55}

\bibitem[{{Ilee} {et~al.}(2021){Ilee}, {Walsh}, {Booth}, {Aikawa}, {Andrews},
  {Bae}, {Bergin}, {Bergner}, {Bosman}, {Cataldi}, {Cleeves}, {Czekala},
  {Guzm{\'a}n}, {Huang}, {Law}, {Le Gal}, {Loomis}, {M{\'e}nard}, {Nomura},
  {{\"O}berg}, {Qi}, {Schwarz}, {Teague}, {Tsukagoshi}, {Wilner}, {Yamato}, \&
  {Zhang}}]{2021ApJS..257....9I}
{Ilee}, J.~D., {Walsh}, C., {Booth}, A.~S., {et~al.} 2021, \apjs, 257, 9,
  \dodoi{10.3847/1538-4365/ac1441}

\bibitem[{{Jeong} {et~al.}(2025){Jeong}, {Lee}, {Lee}, {Baek}, {Kang}, {Lee},
  {Kim}, {Yun}, {Aikawa}, {Herczeg}, {Johnstone}, \&
  {Cieza}}]{2025ApJS..276...49J}
{Jeong}, J.-H., {Lee}, J.-E., {Lee}, S., {et~al.} 2025, \apjs, 276, 49,
  \dodoi{10.3847/1538-4365/ad9450}

\bibitem[{{J{\o}rgensen} {et~al.}(2020){J{\o}rgensen}, {Belloche}, \&
  {Garrod}}]{2020ARA&A..58..727J}
{J{\o}rgensen}, J.~K., {Belloche}, A., \& {Garrod}, R.~T. 2020, \araa, 58, 727,
  \dodoi{10.1146/annurev-astro-032620-021927}

\bibitem[{{J{\o}rgensen} {et~al.}(2016){J{\o}rgensen}, {van der Wiel},
  {Coutens}, {Lykke}, {M{\"u}ller}, {van Dishoeck}, {Calcutt}, {Bjerkeli},
  {Bourke}, {Drozdovskaya}, {Favre}, {Fayolle}, {Garrod}, {Jacobsen},
  {{\"O}berg}, {Persson}, \& {Wampfler}}]{2016A&A...595A.117J}
{J{\o}rgensen}, J.~K., {van der Wiel}, M.~H.~D., {Coutens}, A., {et~al.} 2016,
  \aap, 595, A117, \dodoi{10.1051/0004-6361/201628648}

\bibitem[{{J{\o}rgensen} {et~al.}(2018){J{\o}rgensen}, {M{\"u}ller}, {Calcutt},
  {Coutens}, {Drozdovskaya}, {{\"O}berg}, {Persson}, {Taquet}, {van Dishoeck},
  \& {Wampfler}}]{2018A&A...620A.170J}
{J{\o}rgensen}, J.~K., {M{\"u}ller}, H.~S.~P., {Calcutt}, H., {et~al.} 2018,
  \aap, 620, A170, \dodoi{10.1051/0004-6361/201731667}

\bibitem[{Kluyver {et~al.}(2016)Kluyver, Ragan-Kelley, P{\'e}rez, Granger,
  Bussonnier, Frederic, Kelley, Hamrick, Grout, Corlay, Ivanov, Avila, Abdalla,
  \& Willing}]{Jupyter}
Kluyver, T., Ragan-Kelley, B., P{\'e}rez, F., {et~al.} 2016, in Positioning and
  Power in Academic Publishing: Players, Agents and Agendas, ed. F.~Loizides \&
  B.~Schmidt, IOS Press, 87 -- 90

\bibitem[{{Langer} {et~al.}(1984){Langer}, {Graedel}, {Frerking}, \&
  {Armentrout}}]{1984ApJ...277..581L}
{Langer}, W.~D., {Graedel}, T.~E., {Frerking}, M.~A., \& {Armentrout}, P.~B.
  1984, \apj, 277, 581, \dodoi{10.1086/161730}

\bibitem[{{Lee} {et~al.}(2019){Lee}, {Codella}, {Li}, \&
  {Liu}}]{2019ApJ...876...63L}
{Lee}, C.-F., {Codella}, C., {Li}, Z.-Y., \& {Liu}, S.-Y. 2019, \apj, 876, 63,
  \dodoi{10.3847/1538-4357/ab15db}

\bibitem[{{Lee} {et~al.}(2024){Lee}, {Nomura}, \&
  {Furuya}}]{2024ApJ...969...41L}
{Lee}, S., {Nomura}, H., \& {Furuya}, K. 2024, \apj, 969, 41,
  \dodoi{10.3847/1538-4357/ad39e3}

\bibitem[{{Lin} {et~al.}(2023){Lin}, {Spezzano}, \&
  {Caselli}}]{2023A&A...669L...6L}
{Lin}, Y., {Spezzano}, S., \& {Caselli}, P. 2023, \aap, 669, L6,
  \dodoi{10.1051/0004-6361/202245524}

\bibitem[{{Loomis} {et~al.}(2018{\natexlab{a}}){Loomis}, {Cleeves},
  {{\"O}berg}, {Aikawa}, {Bergner}, {Furuya}, {Guzman}, \&
  {Walsh}}]{2018ApJ...859..131L}
{Loomis}, R.~A., {Cleeves}, L.~I., {{\"O}berg}, K.~I., {et~al.}
  2018{\natexlab{a}}, \apj, 859, 131, \dodoi{10.3847/1538-4357/aac169}

\bibitem[{{Loomis} {et~al.}(2018{\natexlab{b}}){Loomis}, {{\"O}berg},
  {Andrews}, {Walsh}, {Czekala}, {Huang}, \& {Rosenfeld}}]{Loomis2018}
{Loomis}, R.~A., {{\"O}berg}, K.~I., {Andrews}, S.~M., {et~al.}
  2018{\natexlab{b}}, \aj, 155, 182, \dodoi{10.3847/1538-3881/aab604}

\bibitem[{{Manigand} {et~al.}(2020){Manigand}, {J{\o}rgensen}, {Calcutt},
  {M{\"u}ller}, {Ligterink}, {Coutens}, {Drozdovskaya}, {van Dishoeck}, \&
  {Wampfler}}]{2020A&A...635A..48M}
{Manigand}, S., {J{\o}rgensen}, J.~K., {Calcutt}, H., {et~al.} 2020, \aap, 635,
  A48, \dodoi{10.1051/0004-6361/201936299}

\bibitem[{{McClure} {et~al.}(2023){McClure}, {Rocha}, {Pontoppidan}, {Crouzet},
  {Chu}, {Dartois}, {Lamberts}, {Noble}, {Pendleton}, {Perotti}, {Qasim},
  {Rachid}, {Smith}, {Sun}, {Beck}, {Boogert}, {Brown}, {Caselli}, {Charnley},
  {Cuppen}, {Dickinson}, {Drozdovskaya}, {Egami}, {Erkal}, {Fraser}, {Garrod},
  {Harsono}, {Ioppolo}, {Jim{\'e}nez-Serra}, {Jin}, {J{\o}rgensen},
  {Kristensen}, {Lis}, {McCoustra}, {McGuire}, {Melnick}, {{\~A}-berg},
  {Palumbo}, {Shimonishi}, {Sturm}, {van Dishoeck}, \&
  {Linnartz}}]{2023NatAs...7..431M}
{McClure}, M.~K., {Rocha}, W.~R.~M., {Pontoppidan}, K.~M., {et~al.} 2023,
  Nature Astronomy, 7, 431, \dodoi{10.1038/s41550-022-01875-w}

\bibitem[{{McMullin} {et~al.}(2007){McMullin}, {Waters}, {Schiebel}, {Young},
  \& {Golap}}]{2007ASPC..376..127M}
{McMullin}, J.~P., {Waters}, B., {Schiebel}, D., {Young}, W., \& {Golap}, K.
  2007, in Astronomical Society of the Pacific Conference Series, Vol. 376,
  Astronomical Data Analysis Software and Systems XVI, ed. R.~A. {Shaw},
  F.~{Hill}, \& D.~J. {Bell}, 127

\bibitem[{{Minissale} {et~al.}(2022){Minissale}, {Aikawa}, {Bergin}, {Bertin},
  {Brown}, {Cazaux}, {Charnley}, {Coutens}, {Cuppen}, {Guzman}, {Linnartz},
  {McCoustra}, {Rimola}, {Schrauwen}, {Toubin}, {Ugliengo}, {Watanabe},
  {Wakelam}, \& {Dulieu}}]{2022ESC.....6..597M}
{Minissale}, M., {Aikawa}, Y., {Bergin}, E., {et~al.} 2022, ACS Earth and Space
  Chemistry, 6, 597, \dodoi{10.1021/acsearthspacechem.1c00357}

\bibitem[{{M{\"u}ller} {et~al.}(2005){M{\"u}ller}, {Schl{\"o}der}, {Stutzki},
  \& {Winnewisser}}]{2005JMoSt.742..215M}
{M{\"u}ller}, H. S.~P., {Schl{\"o}der}, F., {Stutzki}, J., \& {Winnewisser}, G.
  2005, Journal of Molecular Structure, 742, 215,
  \dodoi{10.1016/j.molstruc.2005.01.027}

\bibitem[{{M{\"u}ller} {et~al.}(2001){M{\"u}ller}, {Thorwirth}, {Roth}, \&
  {Winnewisser}}]{2001A&A...370L..49M}
{M{\"u}ller}, H.~S.~P., {Thorwirth}, S., {Roth}, D.~A., \& {Winnewisser}, G.
  2001, \aap, 370, L49, \dodoi{10.1051/0004-6361:20010367}

\bibitem[{{Nazari} {et~al.}(2021){Nazari}, {van Gelder}, {van Dishoeck},
  {Tabone}, {van't Hoff}, {Ligterink}, {Beuther}, {Boogert}, {Caratti o
  Garatti}, {Klaassen}, {Linnartz}, {Taquet}, \&
  {Tychoniec}}]{2021A&A...650A.150N}
{Nazari}, P., {van Gelder}, M.~L., {van Dishoeck}, E.~F., {et~al.} 2021, \aap,
  650, A150, \dodoi{10.1051/0004-6361/202039996}

\bibitem[{{Notsu} {et~al.}(2019){Notsu}, {Akiyama}, {Booth}, {Nomura}, {Walsh},
  {Hirota}, {Honda}, {Tsukagoshi}, \& {Millar}}]{2019ApJ...875...96N}
{Notsu}, S., {Akiyama}, E., {Booth}, A., {et~al.} 2019, \apj, 875, 96,
  \dodoi{10.3847/1538-4357/ab0ae9}

\bibitem[{{{\"O}berg} {et~al.}(2009){{\"O}berg}, {Garrod}, {van Dishoeck}, \&
  {Linnartz}}]{2009A&A...504..891O}
{{\"O}berg}, K.~I., {Garrod}, R.~T., {van Dishoeck}, E.~F., \& {Linnartz}, H.
  2009, \aap, 504, 891, \dodoi{10.1051/0004-6361/200912559}

\bibitem[{{{\"O}berg} {et~al.}(2015){{\"O}berg}, {Guzm{\'a}n}, {Furuya}, {Qi},
  {Aikawa}, {Andrews}, {Loomis}, \& {Wilner}}]{2015Natur.520..198O}
{{\"O}berg}, K.~I., {Guzm{\'a}n}, V.~V., {Furuya}, K., {et~al.} 2015, \nat,
  520, 198, \dodoi{10.1038/nature14276}

\bibitem[{{Pegues} {et~al.}(2020){Pegues}, {{\"O}berg}, {Bergner}, {Loomis},
  {Qi}, {Le Gal}, {Cleeves}, {Guzm{\'a}n}, {Huang}, {J{\o}rgensen}, {Andrews},
  {Blake}, {Carpenter}, {Schwarz}, {Williams}, \&
  {Wilner}}]{2020ApJ...890..142P}
{Pegues}, J., {{\"O}berg}, K.~I., {Bergner}, J.~B., {et~al.} 2020, \apj, 890,
  142, \dodoi{10.3847/1538-4357/ab64d9}

\bibitem[{P\'erez \& Granger(2007)}]{IPython}
P\'erez, F., \& Granger, B.~E. 2007, Computing in Science and Engineering, 9,
  21, \dodoi{10.1109/MCSE.2007.53}

\bibitem[{{Pickett} {et~al.}(1998){Pickett}, {Poynter}, {Cohen}, {Delitsky},
  {Pearson}, \& {M{\"u}ller}}]{1998JQSRT..60..883P}
{Pickett}, H.~M., {Poynter}, R.~L., {Cohen}, E.~A., {et~al.} 1998, \jqsrt, 60,
  883, \dodoi{10.1016/S0022-4073(98)00091-0}

\bibitem[{{Rosotti} {et~al.}(2020){Rosotti}, {Benisty}, {Juh{\'a}sz}, {Teague},
  {Clarke}, {Dominik}, {Dullemond}, {Klaassen}, {Matr{\`a}}, \&
  {Stolker}}]{2020MNRAS.491.1335R}
{Rosotti}, G.~P., {Benisty}, M., {Juh{\'a}sz}, A., {et~al.} 2020, \mnras, 491,
  1335, \dodoi{10.1093/mnras/stz3090}

\bibitem[{{Rubin} {et~al.}(2019){Rubin}, {Altwegg}, {Balsiger}, {Berthelier},
  {Combi}, {De Keyser}, {Drozdovskaya}, {Fiethe}, {Fuselier}, {Gasc},
  {Gombosi}, {H{\"a}nni}, {Hansen}, {Mall}, {R{\`e}me}, {Schroeder},
  {Schuhmann}, {S{\'e}mon}, {Waite}, {Wampfler}, \&
  {Wurz}}]{2019MNRAS.489..594R}
{Rubin}, M., {Altwegg}, K., {Balsiger}, H., {et~al.} 2019, \mnras, 489, 594,
  \dodoi{10.1093/mnras/stz2086}

\bibitem[{{Scibelli} {et~al.}(2021){Scibelli}, {Shirley}, {Vasyunin}, \&
  {Launhardt}}]{2021MNRAS.504.5754S}
{Scibelli}, S., {Shirley}, Y., {Vasyunin}, A., \& {Launhardt}, R. 2021, \mnras,
  504, 5754, \dodoi{10.1093/mnras/stab1151}

\bibitem[{{Smirnov-Pinchukov} {et~al.}(2022){Smirnov-Pinchukov}, {Mo{\'o}r},
  {Semenov}, {{\'A}brah{\'a}m}, {Henning}, {K{\'o}sp{\'a}l}, {Hughes}, \& {di
  Folco}}]{2022MNRAS.510.1148S}
{Smirnov-Pinchukov}, G.~V., {Mo{\'o}r}, A., {Semenov}, D.~A., {et~al.} 2022,
  \mnras, 510, 1148, \dodoi{10.1093/mnras/stab3146}

\bibitem[{{Smith} \& {Adams}(1980)}]{1980ApJ...242..424S}
{Smith}, D., \& {Adams}, N.~G. 1980, \apj, 242, 424, \dodoi{10.1086/158476}

\bibitem[{{Stapper} {et~al.}(2024){Stapper}, {Hogerheijde}, {van Dishoeck},
  {Lin}, {Ahmadi}, {Booth}, {Grant}, {Immer}, {Leemker}, \&
  {P{\'e}rez-S{\'a}nchez}}]{2024A&A...682A.149S}
{Stapper}, L.~M., {Hogerheijde}, M.~R., {van Dishoeck}, E.~F., {et~al.} 2024,
  \aap, 682, A149, \dodoi{10.1051/0004-6361/202347271}

\bibitem[{{Taquet} {et~al.}(2015){Taquet}, {L{\'o}pez-Sepulcre}, {Ceccarelli},
  {Neri}, {Kahane}, \& {Charnley}}]{2015ApJ...804...81T}
{Taquet}, V., {L{\'o}pez-Sepulcre}, A., {Ceccarelli}, C., {et~al.} 2015, \apj,
  804, 81, \dodoi{10.1088/0004-637X/804/2/81}

\bibitem[{{Temmink} {et~al.}(2023){Temmink}, {Booth}, {van der Marel}, \& {van
  Dishoeck}}]{2023A&A...675A.131T}
{Temmink}, M., {Booth}, A.~S., {van der Marel}, N., \& {van Dishoeck}, E.~F.
  2023, \aap, 675, A131, \dodoi{10.1051/0004-6361/202346272}

\bibitem[{{Temmink} {et~al.}(2024){Temmink}, {Booth}, {Leemker}, {van der
  Marel}, {van Dishoeck}, {Evans}, {Keyte}, {Law}, {Notsu}, {{\"O}berg}, \&
  {Walsh}}]{2024arXiv241112418T}
{Temmink}, M., {Booth}, A.~S., {Leemker}, M., {et~al.} 2024, arXiv e-prints,
  arXiv:2411.12418, \dodoi{10.48550/arXiv.2411.12418}

\bibitem[{{Tobin} {et~al.}(2023){Tobin}, {van't Hoff}, {Leemker}, {van
  Dishoeck}, {Paneque-Carre{\~n}o}, {Furuya}, {Harsono}, {Persson}, {Cleeves},
  {Sheehan}, \& {Cieza}}]{2023Natur.615..227T}
{Tobin}, J.~J., {van't Hoff}, M. L.~R., {Leemker}, M., {et~al.} 2023, \nat,
  615, 227, \dodoi{10.1038/s41586-022-05676-z}

\bibitem[{{van der Marel} {et~al.}(2021){van der Marel}, {Booth}, {Leemker},
  {van Dishoeck}, \& {Ohashi}}]{2021A&A...651L...5V}
{van der Marel}, N., {Booth}, A.~S., {Leemker}, M., {van Dishoeck}, E.~F., \&
  {Ohashi}, S. 2021, \aap, 651, L5, \dodoi{10.1051/0004-6361/202141051}

\bibitem[{{van der Plas} {et~al.}(2019){van der Plas}, {M{\'e}nard},
  {Gonzalez}, {Perez}, {Rodet}, {Pinte}, {Cieza}, {Casassus}, \&
  {Benisty}}]{2019A&A...624A..33V}
{van der Plas}, G., {M{\'e}nard}, F., {Gonzalez}, J.~F., {et~al.} 2019, \aap,
  624, A33, \dodoi{10.1051/0004-6361/201834134}

\bibitem[{{van Gelder} {et~al.}(2020){van Gelder}, {Tabone}, {Tychoniec}, {van
  Dishoeck}, {Beuther}, {Boogert}, {Caratti o Garatti}, {Klaassen}, {Linnartz},
  {M{\"u}ller}, \& {Taquet}}]{2020A&A...639A..87V}
{van Gelder}, M.~L., {Tabone}, B., {Tychoniec}, {\L}., {et~al.} 2020, \aap,
  639, A87, \dodoi{10.1051/0004-6361/202037758}

\bibitem[{{van Gelder} {et~al.}(2022){van Gelder}, {Jaspers}, {Nazari},
  {Ahmadi}, {van Dishoeck}, {Beltr{\'a}n}, {Fuller}, {S{\'a}nchez-Monge}, \&
  {Schilke}}]{2022A&A...667A.136V}
{van Gelder}, M.~L., {Jaspers}, J., {Nazari}, P., {et~al.} 2022, \aap, 667,
  A136, \dodoi{10.1051/0004-6361/202244471}

\bibitem[{{Wagner} {et~al.}(2015){Wagner}, {Apai}, {Kasper}, \&
  {Robberto}}]{2015ApJ...813L...2W}
{Wagner}, K., {Apai}, D., {Kasper}, M., \& {Robberto}, M. 2015, \apjl, 813, L2,
  \dodoi{10.1088/2041-8205/813/1/L2}

\bibitem[{{Wagner} {et~al.}(2018){Wagner}, {Dong}, {Sheehan}, {Apai}, {Kasper},
  {McClure}, {Morzinski}, {Close}, {Males}, {Hinz}, {Quanz}, \&
  {Fung}}]{2018ApJ...854..130W}
{Wagner}, K., {Dong}, R., {Sheehan}, P., {et~al.} 2018, \apj, 854, 130,
  \dodoi{10.3847/1538-4357/aaa767}

\bibitem[{{Walsh} {et~al.}(2014){Walsh}, {Herbst}, {Nomura}, {Millar}, \&
  {Weaver}}]{2014FaDi..168..389W}
{Walsh}, C., {Herbst}, E., {Nomura}, H., {Millar}, T.~J., \& {Weaver}, S.~W.
  2014, Faraday Discussions, 168, 389, \dodoi{10.1039/C3FD00135K}

\bibitem[{{Walsh} {et~al.}(2018){Walsh}, {Vissapragada}, \&
  {McGee}}]{2018IAUS..332..395W}
{Walsh}, C., {Vissapragada}, S., \& {McGee}, H. 2018, in IAU Symposium, Vol.
  332, IAU Symposium, ed. M.~{Cunningham}, T.~{Millar}, \& Y.~{Aikawa},
  395--402, \dodoi{10.1017/S1743921317007037}

\bibitem[{{Walsh} {et~al.}(2016){Walsh}, {Loomis}, {{\"O}berg}, {Kama}, {van 't
  Hoff}, {Millar}, {Aikawa}, {Herbst}, {Widicus Weaver}, \&
  {Nomura}}]{2016ApJ...823L..10W}
{Walsh}, C., {Loomis}, R.~A., {{\"O}berg}, K.~I., {et~al.} 2016, \apjl, 823,
  L10, \dodoi{10.3847/2041-8205/823/1/L10}

\bibitem[{{Wilson}(1999)}]{1999RPPh...62..143W}
{Wilson}, T.~L. 1999, Reports on Progress in Physics, 62, 143,
  \dodoi{10.1088/0034-4885/62/2/002}

\bibitem[{{W{\"o}lfer} {et~al.}(2023){W{\"o}lfer}, {Facchini}, {van der Marel},
  {van Dishoeck}, {Benisty}, {Bohn}, {Francis}, {Izquierdo}, \&
  {Teague}}]{2023A&A...670A.154W}
{W{\"o}lfer}, L., {Facchini}, S., {van der Marel}, N., {et~al.} 2023, \aap,
  670, A154, \dodoi{10.1051/0004-6361/202243601}

\bibitem[{{Woods} \& {Willacy}(2009)}]{2009ApJ...693.1360W}
{Woods}, P.~M., \& {Willacy}, K. 2009, \apj, 693, 1360,
  \dodoi{10.1088/0004-637X/693/2/1360}

\bibitem[{{Yamato} {et~al.}(2024{\natexlab{a}}){Yamato}, {Notsu}, {Aikawa},
  {Okoda}, {Nomura}, \& {Sakai}}]{2024AJ....167...66Y}
{Yamato}, Y., {Notsu}, S., {Aikawa}, Y., {et~al.} 2024{\natexlab{a}}, \aj, 167,
  66, \dodoi{10.3847/1538-3881/ad11d9}

\bibitem[{{Yamato} {et~al.}(2024{\natexlab{b}}){Yamato}, {Aikawa},
  {Guzm{\'a}n}, {Furuya}, {Notsu}, {Cataldi}, {{\"O}berg}, {Qi}, {Law},
  {Huang}, {Teague}, \& {Le Gal}}]{2024ApJ...974...83Y}
{Yamato}, Y., {Aikawa}, Y., {Guzm{\'a}n}, V.~V., {et~al.} 2024{\natexlab{b}},
  \apj, 974, 83, \dodoi{10.3847/1538-4357/ad6981}

\bibitem[{{Yoshida} {et~al.}(2022{\natexlab{a}}){Yoshida}, {Nomura}, {Furuya},
  {Tsukagoshi}, \& {Lee}}]{2022ApJ...932..126Y}
{Yoshida}, T.~C., {Nomura}, H., {Furuya}, K., {Tsukagoshi}, T., \& {Lee}, S.
  2022{\natexlab{a}}, \apj, 932, 126, \dodoi{10.3847/1538-4357/ac6efb}

\bibitem[{{Yoshida} {et~al.}(2022{\natexlab{b}}){Yoshida}, {Nomura},
  {Tsukagoshi}, {Furuya}, \& {Ueda}}]{2022ApJ...937L..14Y}
{Yoshida}, T.~C., {Nomura}, H., {Tsukagoshi}, T., {Furuya}, K., \& {Ueda}, T.
  2022{\natexlab{b}}, \apjl, 937, L14, \dodoi{10.3847/2041-8213/ac903a}

\bibitem[{{Yoshida} {et~al.}(2024){Yoshida}, {Nomura}, {Furuya}, {Teague},
  {Law}, {Tsukagoshi}, {Lee}, {Rab}, {{\"O}berg}, \&
  {Loomis}}]{2024ApJ...966...63Y}
{Yoshida}, T.~C., {Nomura}, H., {Furuya}, K., {et~al.} 2024, \apj, 966, 63,
  \dodoi{10.3847/1538-4357/ad2fb4}

\bibitem[{{Zhang} {et~al.}(2017){Zhang}, {Bergin}, {Blake}, {Cleeves}, \&
  {Schwarz}}]{2017NatAs...1E.130Z}
{Zhang}, K., {Bergin}, E.~A., {Blake}, G.~A., {Cleeves}, L.~I., \& {Schwarz},
  K.~R. 2017, Nature Astronomy, 1, 0130, \dodoi{10.1038/s41550-017-0130}

\bibitem[{{Zhang} {et~al.}(2021){Zhang}, {Snellen}, {Bohn}, {Molli{\`e}re},
  {Ginski}, {Hoeijmakers}, {Kenworthy}, {Mamajek}, {Meshkat}, {Reggiani}, \&
  {Snik}}]{2021Natur.595..370Z}
{Zhang}, Y., {Snellen}, I. A.~G., {Bohn}, A.~J., {et~al.} 2021, \nat, 595, 370,
  \dodoi{10.1038/s41586-021-03616-x}

\end{thebibliography}
\bibliographystyle{aasjournal}


\end{document}